\documentclass[prl,twocolumn,amsmath,amssymb,floatfix,showpacs,superscriptaddress,nofootinbib]{revtex4-2}
\usepackage{graphicx}
\usepackage{amsmath}
\usepackage{placeins}
\usepackage{bm}
\usepackage[dvipsnames]{xcolor}
\usepackage[caption=false]{subfig}
\usepackage{subcaption}
\usepackage{placeins}
\usepackage{ragged2e}
\usepackage{svg}
\usepackage{float}

\begin{document}

\title{Clustering and emergent hyperuniformity by breaking microswimmer shape and actuation symmetries}
\author{Anson G. Thambi}
\affiliation{Department of Mechanical Engineering, University of Hawai’i at 
M{\=a}noa, 2540 Dole Street, Holmes Hall 302, Honolulu, Hawaii 96822, USA}

\author{William E. Uspal}
\email{uspal@hawaii.edu}
\affiliation{Department of Mechanical Engineering, University of Hawai’i at 
M{\=a}noa, 2540 Dole Street, Holmes Hall 302, Honolulu, Hawaii 96822, USA}
\affiliation{International Institute for Sustainability with Knotted Chiral Meta Matter (WPI-SKCM\textsuperscript{2}),
1-3-1 Kagamiyama, Higashi-Hiroshima, Hiroshima 739-8526, Japan}

\begin{abstract}
Hydrodynamic interactions driven by particle activity are ubiquitous in active colloidal systems. Although these interactions are strongly influenced by the interfacial actuation mechanism and geometry of the swimming particles, theoretical understanding of how these microscopic design parameters govern collective dynamics remains limited. Here, we investigate the collective dynamics of oblate spheroidal microswimmers. Using an approximate kinetic theory and corroborating boundary element method calculations, we demonstrate that breaking symmetries in both particle shape and interfacial actuation enables the emergence of dynamically stable immotile $n$-particle clusters. At larger scales, the clustering process drives the system into a dynamically arrested absorbing state characterized by disordered class I hyperuniform structures. Our analysis highlights the essential role of cluster-sourced long-range flows in establishing this long-range order. Overall, our findings reveal a robust, purely hydrodynamic mechanism for hierarchical self-organization in active matter systems, providing a novel strategy for engineering multifunctional hyperuniform materials.
\end{abstract}

\maketitle

\section{Introduction}

Active matter comprises material systems in which individual particulate components consume energy and exert mechanical forces, including propulsive forces that drive self-motility. Inherently, these systems operate out of thermal equilibrium, and they can exhibit exotic collective phenomena and ordering, such as globally aligned ``flocking'' states \cite{vicsek1995novel,kaiser2017flocking}, active turbulence \cite{kokot2017active}, giant number fluctuations \cite{zhang2010collective}, and active clustering / phase separation \cite{buttinoni2013dynamical,cates2015motility,zhang2021active, chen2023tunable}.  

Active systems hold considerable promise for realization of functional materials that can self-organize and structurally reconfigure upon application and modulation of the particles' energy source. 
Moreover, if the self-organized system can exhibit different types of structural order on different length scales, multiple functionalities can be realized within one system. For instance, hierarchically structured colloidal materials can integrate plasmonic resonance and Bragg diffraction for photonic applications \cite{vogel2015color}. 
Hierarchical self-organization offers a natural paradigm for realization of hierarchically structured active materials \cite{needleman2017active}. Initially, order emerges at short length and time scales, as seen when active colloids form small clusters. Over longer time scales, the small self-organized units further organize into larger-scale structures.

Various kinds of self-organized structural order have been observed in active matter systems, such as liquid \cite{geyer2019freezing}, hexatic \cite{digregorio2018full}, and crystalline \cite{petroff2015fast, briand2016crystallization} phases. Recently, the idea of  hyperuniformity has been developed as a new conceptual paradigm in condensed matter \cite{torquato2018hyperuniform}.  Disordered hyperuniform states are exotic states of matter that, like crystals, are accompanied by anomalous suppression of large-scale density fluctuations, yet they preserve the isotropy and absence of Bragg peaks typical of liquids or glasses. This unusual blend of order and disorder gives rise to novel physical and thermodynamic properties. One of the most significant among these is the ability to support complete and isotropic photonic band gaps comparable to those found in photonic crystals, but with enhanced robustness against structural defects and fabrication imperfections. The lack of long-range periodicity further enables direction-independent waveguiding and the realization of novel optical geometries \cite{yu2021engineered}.

A critical step in harnessing the promising applications of hyperuniform states is their physical realization. In this regard, non-equilibrium systems offer a compelling advantage over their equilibrium counterparts \cite{lei2024non}. While equilibrium systems typically require finely tuned long-range repulsive interactions, such as those found in charged binary plasmas \cite{lomba2018disordered,lomba2017disordered}, to achieve self-organized hyperuniformity, non-equilibrium systems can do so even in the absence of such interactions \cite{torquato2018hyperuniform}. In many cases, hyperuniformity in non-equilibrium settings emerges through short-range dynamics, local feedback mechanisms, and collective organization.\cite{lei2024non}.

That said, long-range interactions also play a critical role in certain classes of non-equilibrium systems, especially in active matter, for the formation of hyperuniform states. Unlike in equilibrium contexts, where engineering long-range interactions can be experimentally challenging, active matter systems naturally generate such interactions through self-propulsion and fluid-mediated coupling (hydrodynamic interactions). Experimental studies exemplify this: Zhang and Snezhko \cite{zhang2022hyperuniform} demonstrated hyperuniformity in chiral colloidal spinners at an oil–water interface, while Huang \textit{et al}. \cite{huang2021circular} observed disordered hyperuniform states in swimming marine algae (\textit{Effrenium voratum}) at the air–liquid interface. In both cases, chirality and hydrodynamic interactions played a crucial role. Chiral motion induces localised circulation, and when the centres of the orbiting motion are considered, they behave diffusively, suppressing long-range transport \cite{lei2019nonequilibrium}. This mechanism allows hyperuniformity to emerge spontaneously, without fine-tuning, in driven active matter systems. A similar phenomenon is observed in vortex systems formed by active rotors on fluid membranes. Here, rotors generate long-range vortex flows that organise the system into a hyperuniform state \cite{oppenheimer2022hyperuniformity}. In these cases, hydrodynamic coupling alone is sufficient to produce global suppression of density fluctuations even in disordered, dynamically evolving systems.

On the other hand, absorbing-state phase transitions offer an alternate route to hyperuniformity in non-equilibrium systems. Absorbing states are states that can be reached, but not left, during the time evolution of the system. Absorbing transitions are often observed as a transition from fluctuating to quiescent behavior \cite{corte2008random}. Theoretical work has connected hyperuniformity with absorbing phase transitions \cite{hexner2015hyperuniformity}.  Models such as the random organisation model \cite{corte2008random}, the Manna model, and their conserved variants \cite{hexner2015hyperuniformity} exhibit hyperuniformity at the critical point of the absorbing phase transition. Studies on periodically sheared suspensions \cite{tjhung2015hyperuniform,corte2008random} and sedimenting colloidal particles \cite{wang2018hyperuniformity} have confirmed this connection between criticality and hyperuniform suppression of fluctuations.

An intriguing direction, in this regard, is to explore the potential of active matter systems, particularly those interacting through long-range hydrodynamic or phoretic forces and capable of self-assembling into dynamic clusters, to serve as platforms for realising hyperuniform states. Such systems may offer a unique opportunity to bridge the mechanisms of self-organised criticality, as seen in absorbing-state transitions, with the spontaneous fluctuation suppression characteristic of steady-state active phases.

Among model active matter systems, interfacially driven microswimmers represent one of the most important classes. These active particles operate in aqueous environments, and their propulsion is due to the generation of fluid flows at the solid/liquid interface. For biological swimmers in this group, the interfacial flow is typically due to the coordinated, time-irreversible motion of a carpet of small (relative to the body size) cilia \cite{ishikawa2024fluid}. On the synthetic side, self-phoretic active colloids generate interfacial flow without any moving parts \cite{moran2017phoretic}. Instead, the interfacial flow is due to the self-generated gradient of a thermodynamic variable (e.g., electrical potential \cite{paxton2006catalytically,kuron2018toward}, temperature \cite{jiang2010active}, or chemical concentration \cite{howse2007self}), in conjunction with a 
 molecular-scale interfacial coupling of the variable and hydrodynamic flow \cite{anderson1989colloid}. In both cases, the hydrodynamic actuation is confined to  a thin interfacial region. This interfacial flow entrains flow in the bulk liquid, leading to directed swimming motion of the particle.  
 
The ``squirmer'' model is a classical fluid mechanical model for interfacially-driven self-propulsion \cite{ishikawa2024fluid}. Originally developed to study quasi-spherical micro-organisms \cite{lighthill1952squirming,Blake71}, it has recently found application to self-phoretic active colloids \cite{popescu2018effective, poehnl2020axisymmetric}. In this model, the interfacial actuation is represented as a prescribed ``slip'' boundary condition for the hydrodynamic flow imposed on the surface of a swimming particle. The collective dynamics of squirmer monolayers has received extensive study. For an unconfined monolayer of spherical squirmers, clustering and intracluster velocity alignment were observed by Ishikawa and Pedley \cite{ishikawa2008coherent}. When spherical squirmers are confined to the vicinity of a confining wall by gravity,  they can form a Wigner fluid through repulsive hydrodynamic flows, in addition to fluctuating chains, clusters, and swarms   \cite{kuhr2019collective}. Other studies considered monolayers of prolate spheroidal squirmers. Particle elongation was observed to enhance alignment and swarm formation, with a dominant role for near-field interactions \cite{kyoya2015shape}. A dense monolayer of rod-like squirmers in a Hele-Shaw cell can exhibit active turbulence and clustering \cite{zantop2022emergent}.  The transient pairing and long-time scattering dynamics of two spherical squirmers has also received consideration  \cite{ishikawa2006hydrodynamic,llopis2010hydrodynamic}. More recent studies have observed that lubrication and other near-contact hydrodynamic interactions  can induce bound states of two spherical squirmers \cite{darveniza2022pairwise} and two prolate squirmers in a narrow slit \cite{theers2016modeling}.

Most studies of interfacially-driven microswimmers, including the studies mentioned above,  have assumed axisymmetric actuation. Recently, effects induced by non-axisymmetric actuation have come into focus. For instance, squirmers with a chiral slip profile can continuously rotate \cite{pak2014generalized,pedley2016squirmers,burada2022hydrodynamics}, form  oscillatory bound states \cite{burada2022hydrodynamics}, and synchronize circular trajectories through hydrodynamic interactions \cite{samatas2023hydrodynamic}. Recently, we demonstrated that an \textit{achiral}, but non-axisymmetric slip profile can lead to formation of immotile, ``head-to-head'' bound pairs of oblate spheroidal squirmers \cite{poehnl2023shape}. Experimentally, head-to-head pairing was observed for metallo-dielectric Janus discs that self-propel in an AC electric field \cite{katuri2022arrested}. In this case, the interfacial mechanism, induced charge electrophoresis, is  electrokinetic in origin, and driven by the AC field \cite{squires2006breaking}. Since the discoidal particles align their semi-major axis with the field direction, the axisymmetry of the particle geometry and surface coverage is broken by the field, and the field induces a non-axisymmetric slip on the metal side. Non-axisymmetric actuation may be important for other active colloids. For instance, for catalytically active particles, non-axisymmetric slip can result from non-axisymmetry in the surface patterning, whether introduced by accident or design \cite{archer2015glancing}, or by the presence of a nearby solid surface that breaks axisymmetry of the particle-sourced concentration field \cite{uspal2015self}. 

From the foregoing, it is clear that non-axisymmetric slip may have a larger role in  self-organization of active colloids than previously recognized. This is especially the case for \textit{achiral} slip, as chiral slip can often be inferred from observations of particle rotation (circle swimming). Our results in Ref. \citenum{poehnl2023shape}  suggest two intriguing questions. First, moving beyond consideration of pairs, we ask whether non-axisymmetric slip can lead to the spontaneous formation of multi-particle clusters. Secondly, extending our view to monolayers, we ask whether the tendency to form pairs, and potentially clusters, has implications for collective dynamics. The present work aims to address these two questions. We find that by breaking the interfacial actuation symmetry of the swimmers, they gain access to dynamically stable fixed points that were otherwise inaccessible. These fixed points emerge due to the symmetries in the flow fields generated by the active stresslets, which remain unstable for axisymmetric actuation. By introducing a symmetry-breaking parameter $\delta$, which measures the deviation of the stresslet from its axisymmetric form, we find that swimmers can form not just stable pairs but multi-particle clusters, depending on $\delta$ and the shape of the swimmers. 

Furthermore, we investigate the collective behaviour of these systems as they evolve over time. As the system organizes into clusters, it undergoes a transition into an absorbing state characterized by dynamical arrest, during which the system exhibits the characteristics of a disordered hyperuniform structure of class I. This is particularly noteworthy, as it reveals a novel route for realizing such structures purely through hydrodynamic interactions in active matter systems, where previous studies on active matter systems typically only achieved class III hyperuniformity.  Since hydrodynamic interactions generically occur for active colloids moving in liquid media, our findings  have broad implications for various model systems. Thus, our work reveals that the combined influence of hydrodynamic interactions and absorbing-state dynamics in active matter offers a robust pathway for realizing multifunctional hyperuniform materials, an avenue that could open up new frontiers in materials design, especially for isotropic photonic, acoustic, and transport applications \cite{florescu2009complete,froufe2017band,zhou2019hyperuniform,cheron2022experimental}.

\section{Theory}

In the squirmer model, the finite size of a swimmer is fully resolved, and the hydrodynamic flow exterior to the particle(s) is obtained by solution of governing partial differential equations (PDEs). As a result, this model can become intractably computationally expensive when applied to the study of collective dynamics. Coarse-grained kinetic theories that model hydrodynamic interactions through approximate analytical expressions can make accessible simulations of collective dynamics on large length scales and long time scales. In particular, Saintillan and Shelley developed an approach  based on equations of motion for an elongated, self-propelling particle in a linear or locally linearized flow. The flow induced by the particle was modeled as being due to an active stresslet, i.e., the leading order far-field term \cite{saintillan2007orientational,saintillan2008instabilities}. Within this framework, they were able to recover long-wavelength instabilities and pattern formation.

Here,  building on the Saintillan-Shelley kinetic theory, we develop a minimal model for microswimmers with non-axisymmetric actuation.  Each particle is restricted to swim in the $xy$ plane with direction $\mathbf{\hat{d}}^i$  with  a self-propulsion velocity $U_s^i$. The velocity of particle $i$ can be written as follows:
\begin{equation}
\label{eq:particle_vel}
\textbf{U}^i = U_s^i   \mathbf{\hat{d}}^i + \textbf{u}(\textbf{x}_i),
\end{equation}
where $\textbf{x}_i$ is the position of particle $i$. The term $\textbf{u}(\textbf{x}_i)$ describes the particle advection due to ambient flow at $\textbf{x}_i$ generated by all other particles.

The particles are also allowed to rotate, assuming their orientation vectors are always in the $xy$ plane. For the rotation of each swimmer, we use the Jeffery equation, which is as follows~\cite{saintillan2018rheology},
\begin{equation}
\mathbf{\dot{\hat{d}}}^i = (\bm{\mathcal{I}}-\mathbf{\hat{d}}^i \mathbf{\hat{d}}^i)\cdot (\Gamma_i \textbf{E}(\mathbf{x}_i)+\textbf{W}(\mathbf{x}_i))\cdot\mathbf{\hat{d}}^i.  
\label{eq:jeffery}
\end{equation}
Here, $\bm{\mathcal{I}}$ is the identity tensor, $\textbf{E}(\mathbf{x})$  is the rate of  strain tensor, given by
\[\textbf{E}(\mathbf{x}) = \frac{1}{2}(\nabla \textbf{u}+\nabla\textbf{u}^T),\]
$\textbf{W}(\mathbf{x})$ is the vorticity tensor, given by
\[\textbf{W}(\mathbf{x}) = \frac{1}{2}(\nabla \textbf{u}-\nabla\textbf{u}^T),\]
and $ \Gamma$ is the particle shape parameter, defined as
\[\Gamma = \frac{1-r_e^2}{1+r_e^2},\] where $r_e^{-1}$ is the ratio between the length of the semiaxis of symmetry of the spheroid to the length of the other two semiaxes. Hence $\Gamma = 0$ for a sphere, $\Gamma > 0$ for a prolate spheroid and $\Gamma <  0$ for an oblate spheroid. In the present work, we are only focusing on oblate spheroids, therefore is $\Gamma$ is always negative in our case. The orientation vector $\mathbf{\hat{d}}^i$ is assumed to be aligned with the minor axis of particle $i$.

We use a far-field hydrodynamic approximation to model the swimmer-generated flow. The flow generated by a particle's swimming activity is approximated with the so-called ``active stresslet,'' the leading-order term in an expansion with distance from the particle's center.  The velocity $\textbf{u}(\textbf{x}_i)$ at the position  $\textbf{x}_i$ of particle $i$ is given as follows~\cite{graham2018microhydrodynamics},

\begin{equation}
\begin{aligned}
u_x(\mathbf{x}_i) =  \sum_{j \neq i}^ N \frac{3 (x_j-x_i)}{8\pi\mu r^5} ((\textbf{x}_j-\textbf{x}_i)\cdot\textbf{S}_j\cdot (\textbf{x}_j-\textbf{x}_i))\\
u_y(\mathbf{x}_i) =  \sum_{j \neq i}^ N \frac{3 (y_j-y_i)}{8\pi\mu r^5} ((\textbf{x}_j-\textbf{x}_i)\cdot\textbf{S}_j\cdot (\textbf{x}_j-\textbf{x}_i))
\end{aligned}
\end{equation}
The sums are taken over all other particles, where $N$ is the number of particles. Here, $r_{ij}$ is the distance between the centers of  particles $i$ and $j$, and $x_i$, $y_i$ are components of the position of particle $i$, and $\mu$ is the dynamic viscosity of the fluid.

\begin{figure}[hbt!]
    \centering
    \includegraphics[scale=0.86]{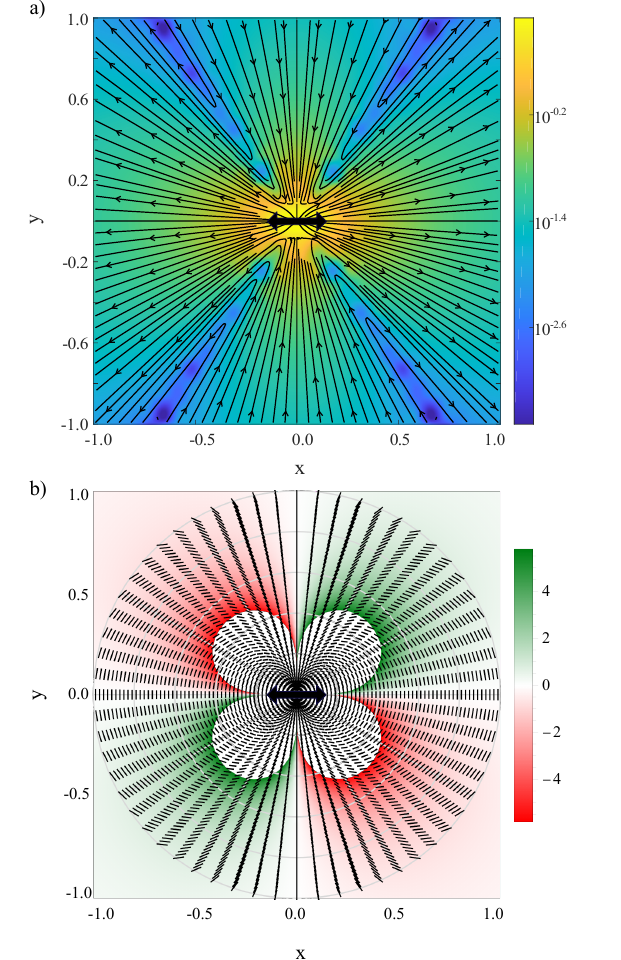}
    \caption{\label{fig:single_stresslet}
    \justifying{(a) Flow field generated by an axisymmetric pusher-type single stresslet ($\sigma_0/\mu L_0^2 U_0 = -6$) located at the origin $(0,0)$. The stresslet, oriented along the $\hat{x}$ direction, is represented by a double-headed arrow. The colormap indicates the magnitude of the velocity field. (b) Vorticity field (background colormap) and elongation axis of the rate-of-strain tensor (black lines) for the same stresslet configuration as in (a). The black lines illustrate the local orientation of the elongation axis throughout the domain. Both the vorticity and rate-of-strain fields exhibit two planes of mirror symmetry.}}
\end{figure}

The quantity $\mathbf{S}_j$, the stresslet tensor, encodes the hydrodynamic properties of particle $j \in \{1, \ldots N\}$. For an axisymmetric squirmer, the stresslet tensor has the following form~\cite{saintillan2018rheology},
\begin{equation}
    \label{axis-stresslet}
    \textbf{S}_0 = \sigma_0 \left(\mathbf{\hat{d}}\mathbf{\hat{d}}-\frac{\bm{\mathcal{I}}}{3}\right).
    \end{equation}
Here, the sign of $\sigma_0$ determines whether the swimmer is a pusher (-ve) or puller (+ve). To impart non-axisymmetry to the stresslet tensor, we rewrite the stresslet tensor in the following form~\cite{poehnl2023shape},
\begin{equation}\
    \label{eq:non-axis}
    \textbf{S} = \textbf{S}_0 + \sigma_0 \delta \left(\mathbf{\hat{c}}\mathbf{\hat{c}}-\mathbf{\hat{e}}\mathbf{\hat{e}}\right).
\end{equation}
Here, $\mathbf{S}$ is written in a reference frame with coordinate axes aligned with the oblate spheroidal particle's axes of geometric symmetry $\mathbf{\hat{c}}$, $\mathbf{\hat{d}}$ and $\mathbf{\hat{e}}$. Recalling that $\mathbf{\hat{d}}$ is aligned with the particle's minor axis, it is clear that $\mathbf{\hat{c}}$ and $\mathbf{\hat{e}}$ are aligned with the two major axes. We assume that $\mathbf{\hat{c}}$ and $\mathbf{\hat{d}}$ lie in the $xy$ plane, and define $\mathbf{\hat{c}} \times \mathbf{\hat{d}} = \mathbf{\hat{e}}$. The dimensionless parameter $\delta$ controls the extent of the deviation from axisymmetry.  When $\delta \neq 0$,  the $\mathbf{\hat{c}} \mathbf{\hat{c}}$ and $\mathbf{\hat{e}} \mathbf{\hat{e}}$ components of $\mathbf{S}$ are no longer identical, i.e., the two major axes of the particle are hydrodynamically distinct. Therefore, we generically expect $\delta \neq 0$ for a particle with non-axisymmetric slip velocity. Note that the form of $\mathbf{S}$ in Eq. \ref{eq:non-axis} satisfies the requirement that $\textrm{tr}(\mathbf{S}) = 0$.

In summary, each particle generates a hydrodynamic flow field based on the above-mentioned non-axisymmetric stresslet. The particles have long-range interactions through these self-generated flows. Hence, using Eqs. \ref{eq:particle_vel} and \ref{eq:jeffery} for the translation and rotation of each particle, we can numerically simulate the motion of $N$ swimmers using the Euler method. Additionally, and as needed, the governing equations can be modified to include a short-ranged excluded volume interaction between particles (see SM Section III) \cite{heyes1993brownian,frenkel2002understanding}. In the following, we assume all particles have the same geometry, self-propulsion speed $U_s$, and stresslet tensor $\mathbf{S}$. For simulation of collective dynamics, we use periodic boundary conditions, as discussed in SM Section III, with box side lengths $L_x = L_y$. The length of the semi-major axis of a particle defines the characteristic length scale $L_0$. The quantity $U_0$ is an arbitrary characteristic velocity. Combining these gives a characteristic time $T_0 \equiv L_0/U_0$. Throughout this manuscript, we use $U_s/U_0 = 0.5$.

\begin{figure*}[hbtp]
    \centering
    \includegraphics[width=\textwidth]{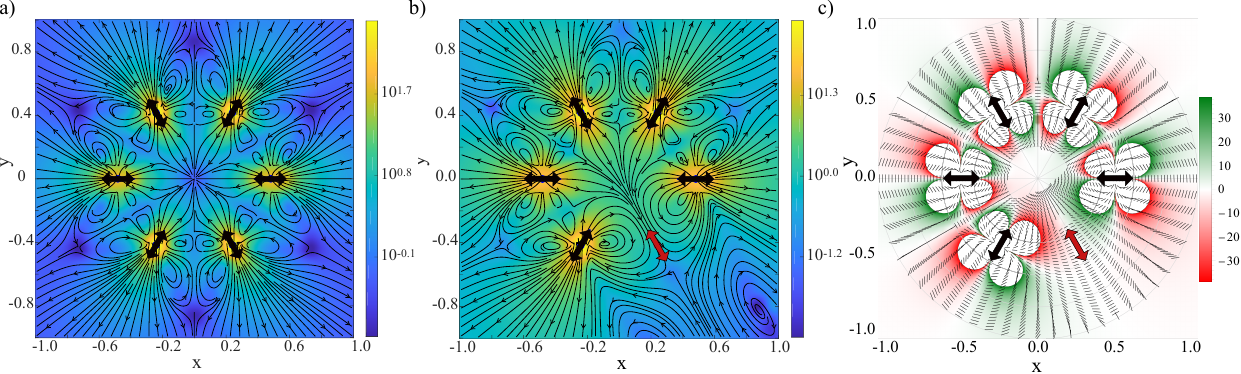}
    \caption{\label{fig:6-particle cluster}\justifying{(a) Flow field generated by six axisymmetric pusher-type stresslets ($\sigma_0/\mu L_0^2 U_0 = -6$) arranged in a hexagonal configuration, with each stresslet oriented (indicated by black arrows) toward the centroid at the origin. (b) Flow field generated by five stresslets in the same hexagonal layout, with the lower-right stresslet removed to reveal the flow field at that location. The position of the missing stresslet is marked by a red double-headed arrow, and the net flow due to the remaining five stresslets is radially outward at that point. (c) Vorticity (colormap) and local elongation axis of the rate-of-strain tensor (black lines) for the configuration in (b). The red arrow again marks the missing stresslet location. The absence of vorticity and the orientation of the elongation axis at this point promote the formation of the hexagonal configuration.}}
\end{figure*}

\section{Results and Discussion}

\subsection{Formation of $n$-particle clusters}

The flow field induced by a stresslet possesses inherent symmetries that imply the existence of dynamical fixed points in the system. In Fig. \ref{fig:single_stresslet}a, flow in the xy plane is shown for an axisymmetric stresslet (Eq. \ref{axis-stresslet}) located at the origin and oriented in the $\mathbf{\hat{d}} = \mathbf{\hat{x}}$ direction.  Fig. \ref{fig:single_stresslet}b shows the vorticity induced by the stresslet and the elongational axis of the rate-of-strain tensor. In both figures, two planes of mirror symmetry are evident. These observations suggest that an arrangement of multiple stresslets, if the arrangement has mirror symmetries, might have a net flow field with the same symmetries.    

Indeed, we find that the symmetries exhibited by an individual stresslet-induced flow field can be retained even when we superpose the flow from multiple stresslets. Fig. \ref{fig:6-particle cluster}a shows the flow field sourced by six axisymmetric stresslets arranged in a hexagonal configuration. The centre of the hexagon is at the origin, and each stresslet has an orientation vector $\mathbf{\hat{d}}$ directed towards the origin. We consider the flow field induced at the position of the bottom right stresslet (indicated by a red arrow) by the other five, as shown in Fig. \ref{fig:6-particle cluster}b. The entire polygonal configuration has a plane of mirror symmetry that runs through the red arrow. As a result of this configurational symmetry and the mirror symmetry of the stresslet-induced flow field, the flow field induced at the position of the bottom right stresslet by the other five stresslets must be in the radial direction. We can see from the Fig. \ref{fig:6-particle cluster}b that this flow is indeed radially outward. If that corresponding stresslet is associated with a self-propelling particle, a cancellation of the particle's net  translational velocity can be realized for some value of the polygon side length $d_n$ (Eq. \ref{eq:particle_vel}). 

Now, we consider the particle's angular velocity. Fig. 2c presents the vorticity of the fluid flow generated by the configuration in Fig. \ref{fig:6-particle cluster}b. Owing to the configurational symmetry of the arrangement, the vorticity at the location of the bottom right stresslet is zero. Now, consider an oblate-shaped particle propelling along the $\mathbf{\hat{d}}$ direction. At this specific location of the bottom right stresslet, the rate of strain steers the orientation vector $\mathbf{\hat{d}}$ inward toward the centre of the polygonal configuration. This behaviour is evident from the extension axis of the rate of the strain tensor, represented by black lines throughout the figure. Notably, at the bottom right stresslet’s position, the extension axis is perpendicular to the line connecting this point to the centroid of the hexagon, forcing the particle's orientation $\mathbf{\hat{d}}$ to be aligned towards the configuration centre. Consequently, this strain pattern promotes the maintenance of the polygonal configuration. Since the particle aligns with the extensional axis of the rate of strain tensor and vorticity is absent at this point, its angular velocity also vanishes. Hence, at the location of the bottom right stresslet, if we associate it with a self-propelling particle with a self-propulsion speed equal to the ambient flow speed, both the translational and angular velocities of the particle are entirely cancelled. This reveals the presence of fixed points within the dynamical system arising from the interplay between the intrinsic mirror symmetry of individual stresslets and configurational symmetry. Notably, since this phenomenon is rooted in the system’s configurational symmetries, such fixed points can be realized for any symmetric polygonal arrangement, regardless of particle number. 
\begin{figure*}[hbtp]
    \centering
    \includegraphics[width=\textwidth]{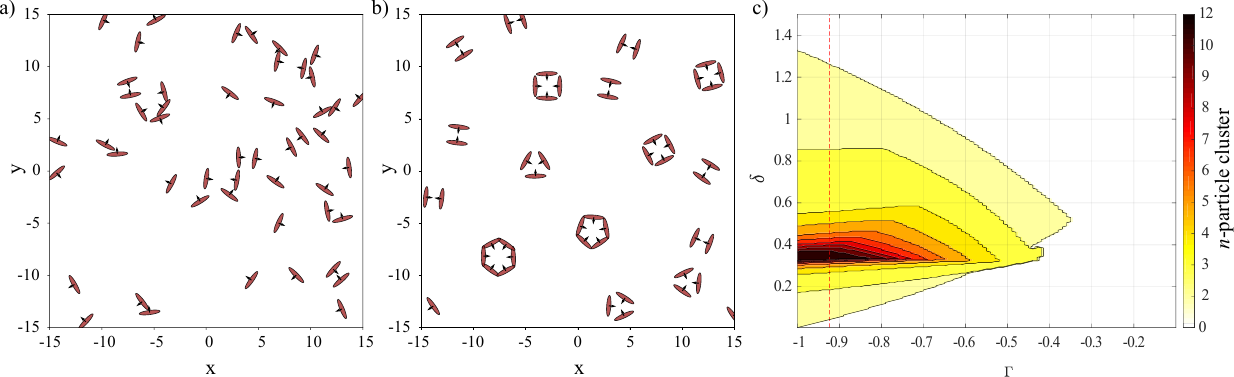}
    \caption{\label{fig:3 cluster}\justifying{(a) Random initial positions and orientations of squirmers at the start of the simulation. (b) Dynamically arrested steady state reached by the squirmers, where particles self-organize into stable $n$-particle clusters. The simulation was performed with parameters $\sigma_0/\mu L_0^2 U_0 = -14.95$, $\delta = 0.333$, and $\Gamma = -0.9231$. (c) Phase map obtained from linear stability analysis using the same stresslet strength, $\sigma_0/\mu L_0^2 U_0 = -14.95$. The red dashed lines indicate the $\Gamma$ value discussed in the collective phenomena section.} In (a) and (b), the axes are shown in units of the semi-major axis length $L_0$, and particles are slightly enlarged for visual clarity.}
\end{figure*}
These fixed points play a crucial role in the self-organization of particles, allowing them to form stable $n$-particle clusters under certain conditions. Even when the stresslet is modified to a non-axisymmetric form by introducing $\delta$ which quantifies the deviation from its axisymmetric state, its inherent symmetry remains preserved. This is evident in the SM Fig. S1, which illustrates the flow field induced by a non-axisymmetric stresslet for various values of $\delta$. Consequently, the introduction of $\delta$ does not disrupt the existence of fixed points, but instead, as will be shown, ensures their stability.
Furthermore, we can find the distance $d_n$ between particles in a regular n-sided polygonal fixed point configuration by setting the velocity $\textbf{U}^i$ to zero. Due to the rotational symmetry of the polygonal arrangement, this equilibrium distance $d_n$ remains identical between all the particle positions in the fixed point configuration.

For an $n$-particle cluster, we can define the angle $\Phi_i$, for each particle in the cluster, which is measured from the x-axis to the centroid of the cluster, as follows

\begin{equation}
    \Phi_i =  \pi + \frac{2\pi(i-1)}{n}
\end{equation}
Then, we can obtain a general expression $d_n$, plotted in the SM Fig. S2, as follows,

\begin{equation}
\label{eq:dnsq}
    \begin{split}
&d_n^2= \frac{\left(-\sigma_0\right) \sin^2\left(\frac{\pi}{n}\right)}{16 \pi U_s \mu \cos\left(\Phi_i\right)} \sum_{j\neq i}^n \frac{1}{\sin^2\frac{\left(i-j\right)\pi}{n}} \biggl[\biggl(1+ 3\delta \; +  \\ &3(\delta-1)\cos\left(\frac{2\pi\left(i+j-2\right)}{n}-2\Phi_j\right)\biggr) \sin\frac{\left(i+j-2\right)\pi}{n}\biggr]
\end{split}
\end{equation}

Having addressed the existence of dynamical fixed points, we now turn to their stability. This is where the parameter $\delta$ plays a role. In  numerical simulations, we find that $\delta \neq 0$ is a necessary condition for particles to spontaneously form $n$-particle clusters from a random initial configuration, as shown in Fig. \ref{fig:3 cluster}. Setting $\delta = 0$ results in no particle cluster formation. These findings are consistent with our earlier work on the formation of bound pairs \cite{poehnl2023shape}. Hence, the $\delta$ parameter can impart stability to the $n$-particle fixed points and is therefore essential for the spontaneous formation of the clusters. We also systematically investigate  how the stability of $n$-particle clusters depends on the value of $\delta$. We use linear stability analysis, as explained in the SM section IV, to obtain a phase diagram of stable states as a function of the  $\delta$ and $\Gamma$ parameters, as shown in Fig. \ref{fig:3 cluster}c.

\begin{figure*}[hbtp]
   \centering
    \includegraphics[scale=.85]{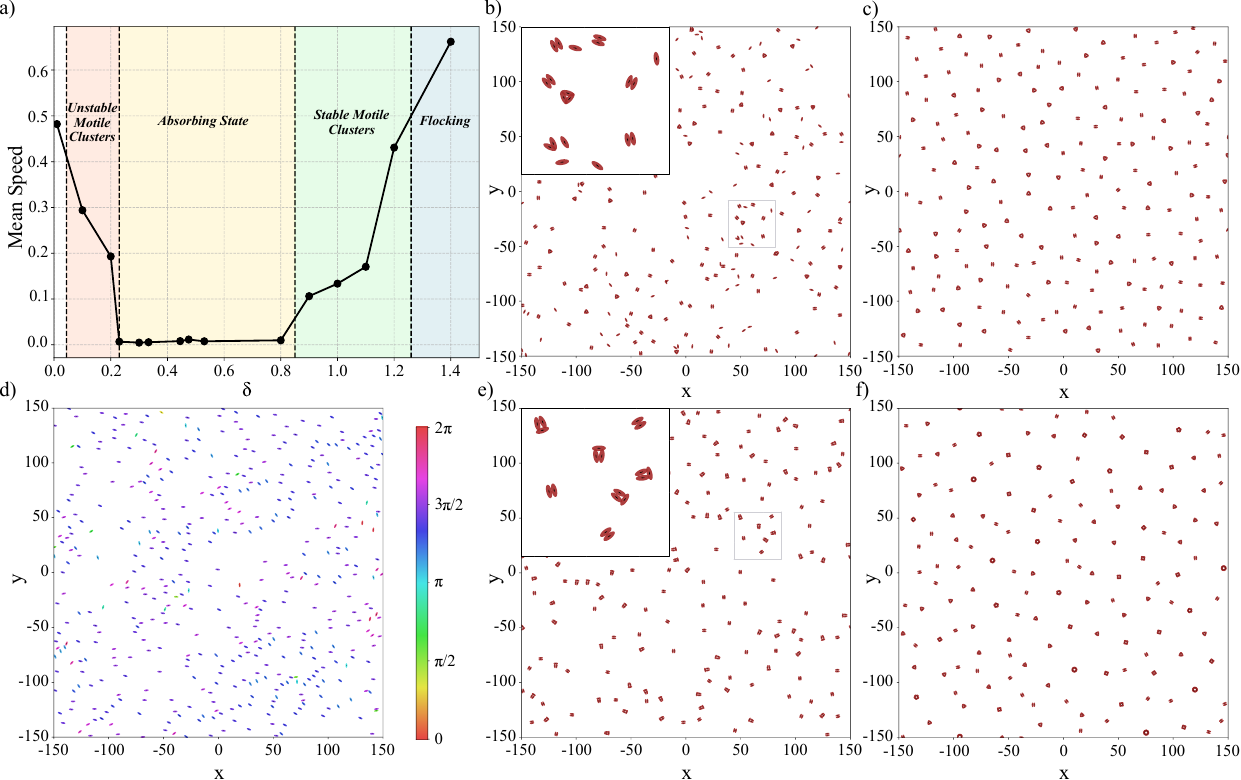}
 \caption{\label{fig:snapshots}\justifying{(a) Mean particle speed (in units of $U_0$) as a function of the asymmetry parameter \(\delta\) after \(10^4\) time units, used to classify the collective dynamical states of the system. The vertical axis shows the average speed of particles, while the horizontal axis corresponds to \(\delta\). Configurations are categorized into four distinct regimes based on whether the mean speed converges to \(\approx 0.01\), indicating dynamical arrest. The shaded regions represent: unstable motile clusters (\(0.04 <\delta \lesssim 0.23\)), absorbing states (\(0.23 \lesssim \delta \le 0.85\)), stable motile clusters (\(0.85 < \delta \le 1.26)\), and flocking behavior (\(\delta > 1.26\)). Snapshots of representative configurations at \(10^4\) time units are shown in panels (b)–(f), using the same parameters: \(\Gamma = -0.9231\), \(\sigma_0/\mu L_0^2 U_0 = -14.95\), and number density \(\phi = 0.005\). (b) Unstable motile clusters at \(\delta = 0.2\). (c) Absorbing state at \(\delta = 0.800\), where particles form up to 3-particle clusters. (d) Flocking state at \(\delta = 1.4\), with particles color-coded by orientation. (e) Stable motile clusters at \(\delta = 1.0\). (f) Absorbing state at \(\delta = 0.333\), where multi-particle clusters are stable. In panels (b) and (e), grey dashed rectangular boxes highlight selected regions that are shown magnified in the respective insets to improve visibility. In all snapshots, the axes are shown in units of the semi-major axis length $L_0$ and the particles are enlarged for visibility (red ellipses); the actual particle size is shown in black within each panel.}}
\end{figure*}

The phase diagram shows that nested stability regions exist in the system. The phase space where stable bound states occur overlaps across different $n$-particle clusters. As the number of particles in an $n$-particle cluster increases, the stable region becomes smaller, and clusters with a higher particle count are nested within the stability regions of clusters with fewer particles. For example, the stability region for a 5-particle cluster lies within that of a 4-particle cluster, which in turn lies within the region for a 3-particle cluster, and so on down to 2-particle clusters. Consequently, the phase map reveals a coexistence region for clusters of varying particle counts. From the phase diagram, we can also conclude that having a high -ve $\Gamma$ is most favourable for them to form clusters. This is when they are maximally deviated from the spherical shape and have a needle-like shape, and hence, it enhances the alignment of the particle's long axis with that of the extension axis of the rate of strain tensor, as per Eq. \ref{eq:jeffery}. We also find that having $\delta = 1/3$ is when maximum $n$-particle clusters are allowed in the system, see Fig. \ref{fig:3 cluster}b and c. As seen from the non-axisymmetric stresslet equation (Eqn. \ref{eq:non-axis}), this particular $\delta$ leads to a complete cancellation of the \(\mathbf{\hat{c}}\mathbf{\hat{c}}\) component of the stresslet, resulting in the elimination of the in-flow component of the stresslet velocity field, as illustrated in SM Fig.~S1 b.
 For $\delta > \frac{1}{3}$ we see that the non-axisymmetric stresslet leads to a purely repulsive flow field in the $xy$ plane (see SM Fig. S1 c and d). The strength of the repulsive behaviour depends on the stresslet strength $\sigma_0$ and self-propulsion velocity $U_s$. Together, they define the proximity of the approach of the particles in the cluster. This can be identified by looking at $d_n$ (Eqn. \ref{eq:dnsq}), where increasing $\sigma_0$ increases the repulsive strength, and hence increases distance between the particles. On the other hand,  increasing the $U_s$ counteracts this repulsive stresslet strength and brings particles close to each other in a cluster. 
 
Our kinetic theory provides only a very approximate treatment of a particle’s finite size. In the squirmer model, the  flow field satisfies a hydrodynamic boundary condition over the entire surface of a particle. In contrast, the kinetic theory accounts for the hydrodynamic boundary condition only in an approximate sense, by using the Jeffery equation to describe the particle's rotation. (This equation describes the rotation of an isolated particle in a linearly varying ambient flow.) Additionally, the kinetic theory does not include near-field hydrodynamic interactions, since the activity-sourced flow field is truncated at the leading order term in the far-field (the stresslet), calculated for an isolated particle. In view of these approximations, we performed numerical calculations with the boundary element method (BEM) to check whether the clusters are stable in a more realistic hydrodynamic model (the squirmer model) \cite{pozrikidis02}. Our implementation of the BEM solves the Stokes equation and associated boundary conditions for a spheroidal squirmer (see SM section V). We find agreement between the squirmer model and the kinetic theory for a range of parameter values of $\Gamma$ and $\delta$  for 3-particle clusters and 4-particle clusters, as shown in the phase diagram Fig. S3.

In the framework of kinetic theory, we also used linear stability analysis to determine for what values of  $\sigma_0$ and $U_s$ the particles could form stable configurations. In our previous work on bound pairs (2-particle clusters), we demonstrated analytically that the stability has no dependence on these parameters \cite{poehnl2023shape}. For clusters with higher particle count, a numerical investigation is needed, owing to the complexity of the governing equations. We find numerically that the stability of $n$-particle clusters is independent of $\sigma_0$ and $U_s$. This means that even when the particles in a regular polygonal configuration are uniformly shifted along the line connecting the centroid to the normal of the edges, the resulting configuration is still a stable fixed point for a correspondingly rescaled $\sigma_0/U_s$ (see the prefactor of Eq. \ref{eq:dnsq}). 

Finally, it is important to emphasize that the particles form clusters without explicit implementation of attractive or repulsive interaction potentials. They swim towards each other due to their activity, and are repelled from each other due to the stresslet associated with them when they have a ``pusher'' character ($\sigma_0 < 0$).

\subsection{Collective Phenomenon}

\begin{figure*}[hbtp]
    \centering
    \includegraphics[width=\textwidth]{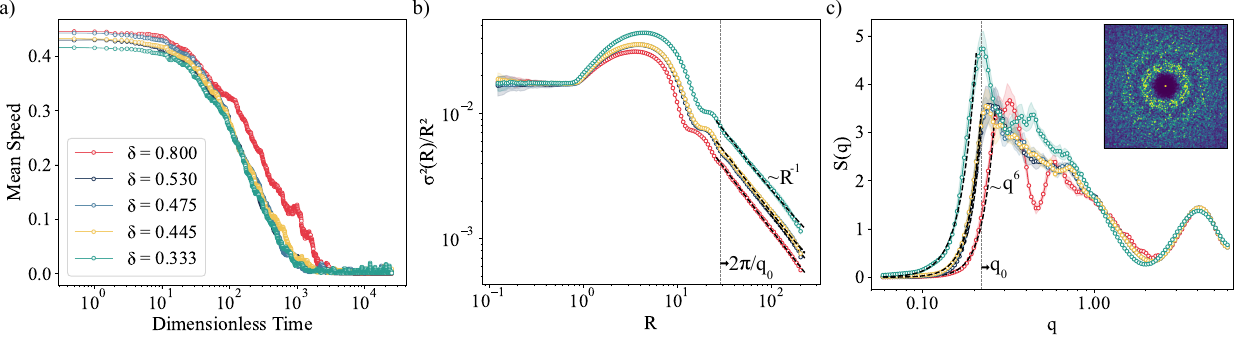}
    \caption{\label{fig:5 hyperuniformity}\justifying{(a) Time evolution of the mean particle speed for representative values of $\delta$ discussed in the hyperuniform structures section. In all cases, the mean speed decays to \(\lesssim 0.01\) within \(10^4\) time units, indicating the emergence of a dynamically arrested absorbing state. Simulations are performed at number density \(\phi = 0.005\) and stresslet strength \(\sigma_0/\mu L_0^2 U_0 = -14.95\).  
(b) Scaled number variance, \(\sigma^2(R)/R^2\), as a function of window radius \(R\) (in units of $L_0$) for the same \(\delta\) values shown in (a). At large \(R\), the variance exhibits a \(1/R\) decay (black dashed line), characteristic of class I hyperuniformity. The vertical grey dashed line marks the crossover length scale \(2\pi/q_0\), where this asymptotic scaling begins.  
(c) Structure factor \(S(q)\) as a function of wavevector \(q\) (in units of $1/L_0$) for the \(\delta\) values listed in (a). The small-\(q\) behavior follows a power-law decay \(S(q) \sim q^6\) (dashed black line), consistent with class I hyperuniformity. The precise power-law exponent extracted from the fit is provided in the SI Table I. The vertical grey dashed line indicates the location of the first peak at \(q_0\), corresponding to the same characteristic length scale identified in (b). The inset shows the two-dimensional structure factor \(S(\mathbf{q})\), confirming isotropy in the $q_x-q_y$ plane.}}
\end{figure*}

Depending on the values of $\delta$ and $\Gamma$, the system can either organize into $n$-particle clusters or remain non-clustering, as illustrated in Fig.~\ref{fig:3 cluster}c. Each scenario gives rise to distinct emergent dynamics on a global scale. To investigate how varying $\delta$ shapes the resulting collective behavior, we fixed $\Gamma = -0.9231$ (red dashed line in the Fig. \ref{fig:3 cluster}c), a value corresponding to $r_e = 5$, and examined the system at a constant number density of $\phi = 0.005$, where $\phi \equiv N L_0^2/L_x^2$. Typical snapshots of the resulting collective states at different $\delta$ values are shown in Fig.~\ref{fig:snapshots}.
Starting from a random initial configuration (Fig. \ref{fig:3 cluster}a), the particles can form \textbf{four} distinct final configurations depending on the value of $\delta$ as shown in Fig. \ref{fig:snapshots}a: \newline
\textbf{1) Flocking.} For $\delta>1.26$, no $n$-particle clusters are formed, but the particles show global polar alignment and hence result in flocking behaviour, as shown in Fig. \ref{fig:snapshots}d and video V1. Fig. S5 shows the evolution of the polar order parameter, defined in the SM section VIII, as a function of time.  \newline
\textbf{2) Stable motile clusters.} When $1.26\ge\delta>0.85$, only $2$-particle immotile clusters are stable. But in this regime, the particles can form stable motile $3$-particle clusters by having a third particle at the rear end of $2$-particle clusters. Since this arrangement deviates from a regular triangular configuration, the net velocity generated by the third particle remains uncanceled, driving the entire configuration forward and hence forming motile $3$-particle clusters. A typical snapshot is shown in Fig. \ref{fig:snapshots}e. These systems remain motile throughout our simulations, as shown in the video V2. \newline
\textbf{3) Absorbing state.} For $0.85\ge\delta\gtrsim0.23$, the system can form stable multi-particle immotile clusters. Typical snapshots for $\delta =  0.800$ and $\delta = 0.3333$, where 3-particle clusters and higher $n$-particle clusters, respectively, are stable are
shown in Fig. \ref{fig:snapshots}c and Fig. \ref{fig:snapshots}f. Corresponding dynamics are provided in Videos V3 and V4. In this regime, the system exhibits a decrease in mean particle speed to approximately zero (i.e., an absorbing transition) and the realization of disordered hyperuniform structure on a global scale. We will discuss these phenomena in detail in the following section. \newline
\textbf{4) Unstable motile clusters.} For $0.23\gtrsim\delta>0.04$, the system still has access to stable fixed points and, therefore, can form multi-particle immotile clusters. We also observe the formation of motile 3-particle clusters. However, most of these cluster formations do not end up in a steady state and, hence, do not show a globally ordered arrangement. A population of single (unclustered) motile particles persists throughout the simulation. Clusters transiently form and, after a finite lifetime, break up through interaction with motile particles.  A typical snapshot of such a configuration is shown in Fig. \ref{fig:snapshots}b, and the dynamical behavior is shown in the video V5.

Our model system thus displays a rich variety of collective behaviours as a function of the parameter $\delta$. In the following, we focus on the absorbing state, which, as we will discuss, exhibits a unique combination of self-organized local and global structural order.

\subsection{Absorbing State and Hyperuniform Structures}

When $\delta$ is between $0.23 $ and $0.85$ for $\Gamma = -0.9231$, starting from a random initial configuration, particles form clusters as time evolves. Since clusters effectively slow down the particles, the system enters an absorbing state, and this can be identified from the average speed of the particles with time, as shown in Fig. \ref{fig:5 hyperuniformity}a. Since within this regime, several n-particle clusters are stable, we look at $\delta$ values $0.8000$, $0.5300$, $0.4750$, $0.4450$, and $0.3333$, which are representative values where the system can form up to 3-particle clusters, 4-particle clusters, 5-particle clusters, 6-particle clusters, and higher $n$-particle clusters, respectively. 

The velocity evolution shows that in all these instances, the system goes into dynamical arrest within $10^4$ time units with a negligible average speed $\approx$ $0.01$, see Fig. \ref{fig:5 hyperuniformity}a. In what follows, we  take this average speed as our criterion for realization of a dynamically arrested absorbing state. 

We now consider the system's large-scale structure in a dynamically arrested state. From Fig.~\ref{fig:snapshots}c and \ref{fig:snapshots}f, it is apparent that the clusters are spread out rather uniformly in space. This appearance is suggestive of a hyperuniform spatial distribution of the clusters. If the clusters were positioned randomly, i.e., if they behaved like an ideal gas, we would expect from ideal gas statistics that some clusters would have close neighbours. Accordingly, we seek to quantitatively confirm the signatures of the hyperuniformity in the system.
One method for identifying hyperuniformity involves examining the local number variance $\sigma^2(R)$. In hyperuniform systems, the variance grows more slowly with $R$ than the window volume, i.e., $\sigma^2(R) \sim R^\beta $, where $d-1 \leq \beta < d$ and $d$ is the system's dimensionality \cite{torquato2018hyperuniform,hawat2023estimating}. Among hyperuniform systems, systems in which $\sigma^2(R)$ grows as $R^{d-1}$ exhibit the strongest form of hyperuniformity and are called class I hyperuniform.  We calculate the variance using the following equation:
\begin{equation}
    \label{variance}
    \sigma^2(R) = \left\langle N^2(R) \right\rangle - \left\langle N(R) \right\rangle^2
\end{equation}
Here, $\left\langle \cdot \right\rangle$ denotes an average over different positions of a window of fixed size $R$ within the simulation box, and $N(R)$ is the number of particles contained in each such window. We also average the values of $\sigma^2(R)$ over ten simulations with different initial conditions. The results of the variance for various $\delta$ values are shown in Fig.~\ref{fig:5 hyperuniformity}b. 

As shown in Fig.~\ref{fig:5 hyperuniformity}b, at very small length scales, the scaled variance $\sigma^2(R)/R^2$ follows the trend expected for a Poisson process. As the window size increases ($R > 1$), the variance exceeds the Poisson level, reaching a peak at intermediate length scales before rapidly falling off and eventually decaying as $1/R$. The intermediate-scale peak reflects the presence of underlying structural features or correlations at that length scale.~\cite{philcox2023disordered}. It arises because observation windows of size comparable to typical particle-particle separations can enclose significantly more or fewer particles depending on whether they overlap with a cluster, thereby enhancing the variance. At larger $R$, these fluctuations average out, and the onset of hyperuniformity leads to the suppression of variance. This behavior reflects a hierarchical structural organization: at small length scales, hydrodynamic interactions among squirmers promote cluster formation, while at larger scales, these clusters are arranged in a manner that yields hyperuniformity. The observed decay of $\sigma^2(R)/R^2$ as $1/R$, as indicated by the $1/R$ fit (shown as dashed black lines in the plot), confirms that the system belongs to class I hyperuniform materials. Notably, the variance curves for each representative $\delta$ value remain fairly consistent, indicating that regardless of which $n$-particle cluster is stable, the system reliably achieves class I hyperuniformity in the absorbing state.

A similar hyperuniform signature is also observed in the static structure factor, $S(\mathbf{q})$, defined as
\begin{equation}
    \label{eq:structure_factor}
    S(\mathbf{q}) =  \frac{1}{N} \left<\left| \sum_{j=1}^{N} \exp\left(-i\mathbf{q} \cdot \mathbf{r}^{(j)}\right) \right|^2\right>,
\end{equation}
where $\mathbf{r}^{(j)}$ denotes the position of the $j$-th particle. Here, the angle brackets $\langle \cdot \rangle$ indicate an ensemble average over ten simulations with different initial conditions. The resulting structure factor $S(q)$ is shown in Fig.~\ref{fig:5 hyperuniformity}c. The inset displays a two-dimensional plot of $S(\mathbf{q})$ in the $q_x$–$q_y$ plane, confirming the isotropy of the system at the level of the structure factor. Details concerning calculation of $S(q)$ are provided in SM Section X \cite{press1992numerical,morris1996spatio,cheng2003fourier,soper2012use}.

As $\mathbf{q} \to 0$, the structure factor exhibits a power-law scaling $S(q) \sim q^6$, as evidenced by the power-law fit (dashed black lines in the plot). This rapid decay is a defining feature of hyperuniform systems. In class I hyperuniform materials, $S(q) \sim q^{\alpha}$ with $\alpha > 1$ in the small-$q$ limit~\cite{torquato2018hyperuniform}; the observed $\alpha \approx 6$ indicates a strong degree of hyperuniformity. This steep scaling suggests that the clusters are arranged in a spatially uniform manner, minimizing long-wavelength density variations. In fact, the static structure factor is strongly attenuated for a range of wavelengths in the vicinity of $q = 0$. This attenutation is strikingly visible as the central dark blue spot in the inset of Fig. \ref{fig:5 hyperuniformity}c. Our system therefore resembles a ``stealthy'' disordered hyperuniform material, with strong suppression of long-wavelength density fluctuations over a finite range of $q$ \cite{florescu2009complete}. However, a more detailed study with larger system sizes would be needed to determine whether our system can be considered stealthy according to strict definition \cite{morse2023generating}. 

Interestingly, the location of the first peak in $S(q)$ (at $q_0$, indicated by the vertical dashed grey line in Fig.~\ref{fig:5 hyperuniformity}c) coincides with the intermediate length scale at which the variance $\sigma^2(R)/R^2$ begins its asymptotic $1/R$ decay (marked by the vertical dashed grey line in Fig.~\ref{fig:5 hyperuniformity}b). This consistency between real-space and Fourier-space diagnostics reinforces the interpretation that both metrics are capturing the same underlying structural organization. Moreover, this behaviour is consistently observed across all representative $\delta$ values, further confirming that the absorbing states reached by the system exhibit class I hyperuniformity regardless of the specific $n$-particle clusters formed.

The emergence of hyperuniform structures has previously been observed in active matter systems, especially in systems of chiral active particles. In such systems, particles undergo persistent circular motion, and their time-averaged hydrodynamic interactions become effectively isotropic and long-ranged. These interactions have a net repulsive effect and lead to hyperuniform configurations, as shown in several recent studies~\cite{huang2021circular,zhang2022hyperuniform,mecke2024emergent}. However, achieving hyperuniformity in achiral active systems, such as linearly swimming microswimmers, is far less straightforward. Unlike chiral particles, whose rotational motion can average out anisotropies, linear swimmers generate highly anisotropic dipolar flow fields. As illustrated in Fig.~\ref{fig:single_stresslet}a, the flow field produced by a pusher stresslet features outward flow in the front and back, and inward flow from the sides. This anisotropy leads to a directional bias in interactions, making it difficult for particles to experience a uniformly repulsive influence at long range. Consequently, the anisotropic nature of these interactions can suppress large-scale order and impede the spontaneous emergence of hyperuniformity.

This is where our study offers fresh insights. To the best of our knowledge, we demonstrate for the first time that hyperuniformity is not exclusive to chiral active particles; achiral microswimmers can also achieve this remarkable state. Although the process is less straightforward than in the chiral case, the underlying mechanism remains similar: emergent repulsion at the level of clusters drives the suppression of long-range density fluctuations. A similar origin of hyperuniformity can be seen in SALR (short-range attractive, long-range repulsive) systems, widely studied in passive matter \cite{imperio2004bidimensional}. These systems exhibit clustering at short length scales due to explicit attractive interactions, while long-range repulsion between clusters leads to suppressed density fluctuations and disordered hyperuniformity \cite{diaz2025build}. Our system, though active and achiral, shows strikingly similar phenomenology. 

In our system, the formation of $n$-particle clusters generates an effective long-range repulsive interaction. Specifically, when particles are organized into clusters, the collective hydrodynamic flow they generate can lead to an emergent repulsive effect at larger scales. For instance, in a regular polygonal arrangement such as the hexagonal configuration illustrated at $\delta = 0$ (see Fig.~\ref{fig:6-particle cluster}a), the resulting flow field is predominantly outward, promoting effective long-range repulsion. Although particles do not spontaneously form stable hexagonal clusters at $\delta = 0$, this example demonstrates that such configurations, if realized, can give rise to outward-directed flows and repulsive hydrodynamic interactions, which are conducive to hyperuniform organization. This suggests that clustering, driven by hydrodynamic interactions, may serve as a general mechanism for enabling hyperuniformity in systems where particles do not possess intrinsic long-range repulsive interactions. Of course, for clusters with small particle number $n$, e.g., $n = 2$, the flow from a cluster cannot be considered effectively radially outward when $\delta = 0$. However, as previously noted, for $\delta \geq \frac{1}{3}$, the flow field from an individual stresslet can itself become purely repulsive (Fig.~S1). This ensures a consistently repulsive interaction between clusters, even when the cluster particle number $n$ is low.

It is also important to note that the emergence of hyperuniformity in the system is consistently accompanied by a dynamically arrested absorbing state. When the system fails to reach this absorbing state, typically for $\delta$ values outside the absorbing regime, the mean particle speed does not decay to the threshold value of $\approx 0.01$ (see Fig.~\ref{fig:5 hyperuniformity}a), and the system does not settle into a steady configuration, as evidenced by the fluctuations in the mean speed over time (see Fig.~S4). This behavior is further reflected in the disordered particle arrangements observed in the snapshots (Fig.~\ref{fig:snapshots}b,d,f). The absence of a well-defined steady state in these cases inhibits the suppression of long-wavelength density fluctuations, and consequently, the system does not exhibit hyperuniformity. These observations suggest that the absorbing state transition, together with hydrodynamic interactions, plays an important role in enabling the development of hyperuniform structures in this system.

\section{Conclusions}

Overall, our study reveals a pathway for hierarchical self-organization of hyperuniform structures for an achiral active system. The particles  self-organize into immotile clusters -- an absorbing state -- and the clusters realize a hyperuniform spatial distribution. Our analysis shows that the hyperuniformity consistently relies on the presence of effective repulsion between clusters. The process does not require any explicit attractive interactions and arises entirely from the interplay of hydrodynamic flows and the system’s evolution toward an absorbing state, offering a new design principle for achieving hyperuniformity in active matter systems.

Furthermore, our model system exhibits a rich variety of collective behaviors as a function of the particle shape and non-axisymmetry. Although our analysis of global structural order focused on hyperuniformity in the absorbing state, future work could  systematically investigate the other observed collective behaviors (unstable motile clusters, stable motile clusters, and flocking). The flocking state, for instance, could be related to previous observations of flocking induced by hydrodynamic interactions for spherical particles \cite{alarcon2013spontaneous,oyama2016purely}. The motile clusters could be studied in microscopic detail with the boundary element method.

Our predictions could potentially be tested experimentally with active particles that self-propel by induced charge electrophoresis. However, additional work is needed to map the design parameters (particle shape, extent of metal coverage, AC field frequency) for ICEP particles to amplitudes of the squirming modes studied here. This mapping can be performed straightforwardly within the framework of the linearized theory for ICEP \cite{squires2006breaking}. 

Future theoretical work could address the related issues of geometric confinement and system size. Specifically, the stresslet flow field considered in this work is the flow field for an unconfined system, and it decays as $\sim 1/r^2$. Additionally, we use the simplest possible realization of periodic boundary conditions (the minimum image convention), and no summation over periodic images is performed. For a 2D monolayer with a $\sim 1/r^2$ interaction, summation of pairwise interactions in an infinite system, or for an infinite series of periodic images, is potentially divergent \cite{bleibel2014hydrodynamic}. On the other hand, the presence of one or two confining planar boundaries would screen the hydrodynamic interaction to have a faster decay with $r$, eliminating any divergence \cite{blake1971note,mitchell2015sedimentation}. A follow-up study could improve on the sophistication of the present approach by incorporating the effects of periodic images \cite{de2017stirring}, as well as confining boundaries via the method of image singularities \cite{mitchell2015sedimentation}. An interesting question is whether a hyperuniform distribution of clusters would still be obtained with a faster decay of the repulsive hydrodynamic interactions between clusters. This question is especially pertinent in view of the fact that most experimental realizations of interfacially-driven active matter involve particles moving in a monolayer near a planar solid surface. 

Our theoretical and numerical framework is easily extended to consideration of heterogeneous mixtures. In previous work, we found that necessary condition for formation of motile ``head-to-tail'' bound pairs is that the particles must have non-identical self-propulsion speeds \cite{poehnl2023shape}.
Moreover, head-to-tail bound pairs are stable against any three-dimensional perturbation, and do not require non-axisymmetric actuation. Work on binary mixtures of prolate and oblate swimmers is currently in progress.

 \section{Acknowledgments} This research was  sponsored by the Army Research Office and was accomplished under Grant Number W911NF-23-1-0190. The views and conclusions contained in this document are those of the authors and should not be interpreted as representing the official policies, either expressed or implied, of the Army Research Office or the U.S. Government. The U.S. Government is authorized to reproduce and distribute reprints for Government purposes notwithstanding any copyright notation herein. The technical support and advanced computing resources from University of Hawaii Information Technology Services – Cyberinfrastructure, funded in part by the National Science Foundation CC* awards \#2201428 and \#2232862 are gratefully acknowledged. This work used Bridges-2 at Pittsburgh Supercomputing Center through allocation PHY250064 from the Advanced Cyberinfrastructure Coordination Ecosystem: Services \& Support (ACCESS) program, which is supported by National Science Foundation grants \#2138259, \#2138286, \#2138307, \#2137603, and \#2138296.

\bibliography{clusters}

\begin{thebibliography}{9}%
\makeatletter
\providecommand \@ifxundefined [1]{%
 \@ifx{#1\undefined}
}%
\providecommand \@ifnum [1]{%
 \ifnum #1\expandafter \@firstoftwo
 \else \expandafter \@secondoftwo
 \fi
}%
\providecommand \@ifx [1]{%
 \ifx #1\expandafter \@firstoftwo
 \else \expandafter \@secondoftwo
 \fi
}%
\providecommand \natexlab [1]{#1}%
\providecommand \enquote  [1]{``#1''}%
\providecommand \bibnamefont  [1]{#1}%
\providecommand \bibfnamefont [1]{#1}%
\providecommand \citenamefont [1]{#1}%
\providecommand \href@noop [0]{\@secondoftwo}%
\providecommand \href [0]{\begingroup \@sanitize@url \@href}%
\providecommand \@href[1]{\@@startlink{#1}\@@href}%
\providecommand \@@href[1]{\endgroup#1\@@endlink}%
\providecommand \@sanitize@url [0]{\catcode `\\12\catcode `\$12\catcode
  `\&12\catcode `\#12\catcode `\^12\catcode `\_12\catcode `\%12\relax}%
\providecommand \@@startlink[1]{}%
\providecommand \@@endlink[0]{}%
\providecommand \url  [0]{\begingroup\@sanitize@url \@url }%
\providecommand \@url [1]{\endgroup\@href {#1}{\urlprefix }}%
\providecommand \urlprefix  [0]{URL }%
\providecommand \Eprint [0]{\href }%
\providecommand \doibase [0]{https://doi.org/}%
\providecommand \selectlanguage [0]{\@gobble}%
\providecommand \bibinfo  [0]{\@secondoftwo}%
\providecommand \bibfield  [0]{\@secondoftwo}%
\providecommand \translation [1]{[#1]}%
\providecommand \BibitemOpen [0]{}%
\providecommand \bibitemStop [0]{}%
\providecommand \bibitemNoStop [0]{.\EOS\space}%
\providecommand \EOS [0]{\spacefactor3000\relax}%
\providecommand \BibitemShut  [1]{\csname bibitem#1\endcsname}%
\let\auto@bib@innerbib\@empty
\bibitem [{\citenamefont {Heyes}\ and\ \citenamefont
  {Melrose}(1993)}]{heyes1993brownian}%
  \BibitemOpen
  \bibfield  {author} {\bibinfo {author} {\bibfnamefont {D.}~\bibnamefont
  {Heyes}}\ and\ \bibinfo {author} {\bibfnamefont {J.}~\bibnamefont
  {Melrose}},\ }\bibfield  {title} {\bibinfo {title} {Brownian dynamics
  simulations of model hard-sphere suspensions},\ }\href@noop {} {\bibfield
  {journal} {\bibinfo  {journal} {Journal of non-newtonian fluid mechanics}\
  }\textbf {\bibinfo {volume} {46}},\ \bibinfo {pages} {1} (\bibinfo {year}
  {1993})}\BibitemShut {NoStop}%
\bibitem [{\citenamefont {Frenkel}\ and\ \citenamefont
  {Smit}(2002)}]{frenkel2002understanding}%
  \BibitemOpen
  \bibfield  {author} {\bibinfo {author} {\bibfnamefont {D.}~\bibnamefont
  {Frenkel}}\ and\ \bibinfo {author} {\bibfnamefont {B.}~\bibnamefont {Smit}},\
  }\href@noop {} {\emph {\bibinfo {title} {Understanding molecular
  simulation}}}\ (\bibinfo  {publisher} {Academic Press},\ \bibinfo {address}
  {San Diego},\ \bibinfo {year} {2002})\BibitemShut {NoStop}%
\bibitem [{\citenamefont {Poehnl}\ \emph {et~al.}(2020)\citenamefont {Poehnl},
  \citenamefont {Popescu},\ and\ \citenamefont
  {Uspal}}]{poehnl2020axisymmetric}%
  \BibitemOpen
  \bibfield  {author} {\bibinfo {author} {\bibfnamefont {R.}~\bibnamefont
  {Poehnl}}, \bibinfo {author} {\bibfnamefont {M.~N.}\ \bibnamefont
  {Popescu}},\ and\ \bibinfo {author} {\bibfnamefont {W.~E.}\ \bibnamefont
  {Uspal}},\ }\bibfield  {title} {\bibinfo {title} {Axisymmetric spheroidal
  squirmers and self-diffusiophoretic particles},\ }\href@noop {} {\bibfield
  {journal} {\bibinfo  {journal} {Journal of Physics: Condensed Matter}\
  }\textbf {\bibinfo {volume} {32}},\ \bibinfo {pages} {164001} (\bibinfo
  {year} {2020})}\BibitemShut {NoStop}%
\bibitem [{\citenamefont {Poehnl}\ and\ \citenamefont
  {Uspal}(2023)}]{poehnl2023shape}%
  \BibitemOpen
  \bibfield  {author} {\bibinfo {author} {\bibfnamefont {R.}~\bibnamefont
  {Poehnl}}\ and\ \bibinfo {author} {\bibfnamefont {W.~E.}\ \bibnamefont
  {Uspal}},\ }\bibfield  {title} {\bibinfo {title} {Shape-induced pairing of
  spheroidal squirmers},\ }\href@noop {} {\bibfield  {journal} {\bibinfo
  {journal} {Physical Review Fluids}\ }\textbf {\bibinfo {volume} {8}},\
  \bibinfo {pages} {113103} (\bibinfo {year} {2023})}\BibitemShut {NoStop}%
\bibitem [{\citenamefont {Hawat}\ \emph {et~al.}(2023)\citenamefont {Hawat},
  \citenamefont {Gautier}, \citenamefont {Bardenet},\ and\ \citenamefont
  {Lachi{\`e}ze-Rey}}]{hawat2023estimating}%
  \BibitemOpen
  \bibfield  {author} {\bibinfo {author} {\bibfnamefont {D.}~\bibnamefont
  {Hawat}}, \bibinfo {author} {\bibfnamefont {G.}~\bibnamefont {Gautier}},
  \bibinfo {author} {\bibfnamefont {R.}~\bibnamefont {Bardenet}},\ and\
  \bibinfo {author} {\bibfnamefont {R.}~\bibnamefont {Lachi{\`e}ze-Rey}},\
  }\bibfield  {title} {\bibinfo {title} {On estimating the structure factor of
  a point process, with applications to hyperuniformity},\ }\href@noop {}
  {\bibfield  {journal} {\bibinfo  {journal} {Statistics and Computing}\
  }\textbf {\bibinfo {volume} {33}},\ \bibinfo {pages} {61} (\bibinfo {year}
  {2023})}\BibitemShut {NoStop}%
\bibitem [{\citenamefont {Soper}\ and\ \citenamefont
  {Barney}(2012)}]{soper2012use}%
  \BibitemOpen
  \bibfield  {author} {\bibinfo {author} {\bibfnamefont {A.~K.}\ \bibnamefont
  {Soper}}\ and\ \bibinfo {author} {\bibfnamefont {E.~R.}\ \bibnamefont
  {Barney}},\ }\bibfield  {title} {\bibinfo {title} {On the use of modification
  functions when fourier transforming total scattering data},\ }\href@noop {}
  {\bibfield  {journal} {\bibinfo  {journal} {Applied Crystallography}\
  }\textbf {\bibinfo {volume} {45}},\ \bibinfo {pages} {1314} (\bibinfo {year}
  {2012})}\BibitemShut {NoStop}%
\bibitem [{\citenamefont {Cheng}\ \emph {et~al.}(2003)\citenamefont {Cheng},
  \citenamefont {Fenter}, \citenamefont {Bedzyk},\ and\ \citenamefont
  {Sturchio}}]{cheng2003fourier}%
  \BibitemOpen
  \bibfield  {author} {\bibinfo {author} {\bibfnamefont {L.}~\bibnamefont
  {Cheng}}, \bibinfo {author} {\bibfnamefont {P.}~\bibnamefont {Fenter}},
  \bibinfo {author} {\bibfnamefont {M.}~\bibnamefont {Bedzyk}},\ and\ \bibinfo
  {author} {\bibfnamefont {N.}~\bibnamefont {Sturchio}},\ }\bibfield  {title}
  {\bibinfo {title} {Fourier-expansion solution of atom distributions in a
  crystal using x-ray standing waves},\ }\href@noop {} {\bibfield  {journal}
  {\bibinfo  {journal} {Physical review letters}\ }\textbf {\bibinfo {volume}
  {90}},\ \bibinfo {pages} {255503} (\bibinfo {year} {2003})}\BibitemShut
  {NoStop}%
\bibitem [{\citenamefont {Press}\ \emph {et~al.}(1992)\citenamefont {Press},
  \citenamefont {Teukolsky}, \citenamefont {Vetterling},\ and\ \citenamefont
  {Flannery}}]{press1992numerical}%
  \BibitemOpen
  \bibfield  {author} {\bibinfo {author} {\bibfnamefont {W.~H.}\ \bibnamefont
  {Press}}, \bibinfo {author} {\bibfnamefont {S.~A.}\ \bibnamefont
  {Teukolsky}}, \bibinfo {author} {\bibfnamefont {W.~T.}\ \bibnamefont
  {Vetterling}},\ and\ \bibinfo {author} {\bibfnamefont {B.~P.}\ \bibnamefont
  {Flannery}},\ }\href@noop {} {\emph {\bibinfo {title} {Numerical Recipes}}}\
  (\bibinfo  {publisher} {Cambridge Univ. Press},\ \bibinfo {year}
  {1992})\BibitemShut {NoStop}%
\bibitem [{\citenamefont {Morris}\ \emph {et~al.}(1996)\citenamefont {Morris},
  \citenamefont {Bodenschatz}, \citenamefont {Cannell},\ and\ \citenamefont
  {Ahlers}}]{morris1996spatio}%
  \BibitemOpen
  \bibfield  {author} {\bibinfo {author} {\bibfnamefont {S.~W.}\ \bibnamefont
  {Morris}}, \bibinfo {author} {\bibfnamefont {E.}~\bibnamefont {Bodenschatz}},
  \bibinfo {author} {\bibfnamefont {D.~S.}\ \bibnamefont {Cannell}},\ and\
  \bibinfo {author} {\bibfnamefont {G.}~\bibnamefont {Ahlers}},\ }\bibfield
  {title} {\bibinfo {title} {The spatio-temporal structure of spiral-defect
  chaos},\ }\href@noop {} {\bibfield  {journal} {\bibinfo  {journal} {Physica
  D: Nonlinear Phenomena}\ }\textbf {\bibinfo {volume} {97}},\ \bibinfo {pages}
  {164} (\bibinfo {year} {1996})}\BibitemShut {NoStop}%
\end{thebibliography}%


\begin{thebibliography}{82}%
\makeatletter
\providecommand \@ifxundefined [1]{%
 \@ifx{#1\undefined}
}%
\providecommand \@ifnum [1]{%
 \ifnum #1\expandafter \@firstoftwo
 \else \expandafter \@secondoftwo
 \fi
}%
\providecommand \@ifx [1]{%
 \ifx #1\expandafter \@firstoftwo
 \else \expandafter \@secondoftwo
 \fi
}%
\providecommand \natexlab [1]{#1}%
\providecommand \enquote  [1]{``#1''}%
\providecommand \bibnamefont  [1]{#1}%
\providecommand \bibfnamefont [1]{#1}%
\providecommand \citenamefont [1]{#1}%
\providecommand \href@noop [0]{\@secondoftwo}%
\providecommand \href [0]{\begingroup \@sanitize@url \@href}%
\providecommand \@href[1]{\@@startlink{#1}\@@href}%
\providecommand \@@href[1]{\endgroup#1\@@endlink}%
\providecommand \@sanitize@url [0]{\catcode `\\12\catcode `\$12\catcode
  `\&12\catcode `\#12\catcode `\^12\catcode `\_12\catcode `\%12\relax}%
\providecommand \@@startlink[1]{}%
\providecommand \@@endlink[0]{}%
\providecommand \url  [0]{\begingroup\@sanitize@url \@url }%
\providecommand \@url [1]{\endgroup\@href {#1}{\urlprefix }}%
\providecommand \urlprefix  [0]{URL }%
\providecommand \Eprint [0]{\href }%
\providecommand \doibase [0]{https://doi.org/}%
\providecommand \selectlanguage [0]{\@gobble}%
\providecommand \bibinfo  [0]{\@secondoftwo}%
\providecommand \bibfield  [0]{\@secondoftwo}%
\providecommand \translation [1]{[#1]}%
\providecommand \BibitemOpen [0]{}%
\providecommand \bibitemStop [0]{}%
\providecommand \bibitemNoStop [0]{.\EOS\space}%
\providecommand \EOS [0]{\spacefactor3000\relax}%
\providecommand \BibitemShut  [1]{\csname bibitem#1\endcsname}%
\let\auto@bib@innerbib\@empty
\bibitem [{\citenamefont {Vicsek}\ \emph {et~al.}(1995)\citenamefont {Vicsek},
  \citenamefont {Czir{\'o}k}, \citenamefont {Ben-Jacob}, \citenamefont
  {Cohen},\ and\ \citenamefont {Shochet}}]{vicsek1995novel}%
  \BibitemOpen
  \bibfield  {author} {\bibinfo {author} {\bibfnamefont {T.}~\bibnamefont
  {Vicsek}}, \bibinfo {author} {\bibfnamefont {A.}~\bibnamefont {Czir{\'o}k}},
  \bibinfo {author} {\bibfnamefont {E.}~\bibnamefont {Ben-Jacob}}, \bibinfo
  {author} {\bibfnamefont {I.}~\bibnamefont {Cohen}},\ and\ \bibinfo {author}
  {\bibfnamefont {O.}~\bibnamefont {Shochet}},\ }\bibfield  {title} {\bibinfo
  {title} {Novel type of phase transition in a system of self-driven
  particles},\ }\href@noop {} {\bibfield  {journal} {\bibinfo  {journal}
  {Physical Review Letters}\ }\textbf {\bibinfo {volume} {75}},\ \bibinfo
  {pages} {1226} (\bibinfo {year} {1995})}\BibitemShut {NoStop}%
\bibitem [{\citenamefont {Kaiser}\ \emph {et~al.}(2017)\citenamefont {Kaiser},
  \citenamefont {Snezhko},\ and\ \citenamefont {Aranson}}]{kaiser2017flocking}%
  \BibitemOpen
  \bibfield  {author} {\bibinfo {author} {\bibfnamefont {A.}~\bibnamefont
  {Kaiser}}, \bibinfo {author} {\bibfnamefont {A.}~\bibnamefont {Snezhko}},\
  and\ \bibinfo {author} {\bibfnamefont {I.~S.}\ \bibnamefont {Aranson}},\
  }\bibfield  {title} {\bibinfo {title} {Flocking ferromagnetic colloids},\
  }\href@noop {} {\bibfield  {journal} {\bibinfo  {journal} {Science Advances}\
  }\textbf {\bibinfo {volume} {3}},\ \bibinfo {pages} {e1601469} (\bibinfo
  {year} {2017})}\BibitemShut {NoStop}%
\bibitem [{\citenamefont {Kokot}\ \emph {et~al.}(2017)\citenamefont {Kokot},
  \citenamefont {Das}, \citenamefont {Winkler}, \citenamefont {Gompper},
  \citenamefont {Aranson},\ and\ \citenamefont {Snezhko}}]{kokot2017active}%
  \BibitemOpen
  \bibfield  {author} {\bibinfo {author} {\bibfnamefont {G.}~\bibnamefont
  {Kokot}}, \bibinfo {author} {\bibfnamefont {S.}~\bibnamefont {Das}}, \bibinfo
  {author} {\bibfnamefont {R.~G.}\ \bibnamefont {Winkler}}, \bibinfo {author}
  {\bibfnamefont {G.}~\bibnamefont {Gompper}}, \bibinfo {author} {\bibfnamefont
  {I.~S.}\ \bibnamefont {Aranson}},\ and\ \bibinfo {author} {\bibfnamefont
  {A.}~\bibnamefont {Snezhko}},\ }\bibfield  {title} {\bibinfo {title} {Active
  turbulence in a gas of self-assembled spinners},\ }\href@noop {} {\bibfield
  {journal} {\bibinfo  {journal} {Proceedings of the National Academy of
  Sciences}\ }\textbf {\bibinfo {volume} {114}},\ \bibinfo {pages} {12870}
  (\bibinfo {year} {2017})}\BibitemShut {NoStop}%
\bibitem [{\citenamefont {Zhang}\ \emph {et~al.}(2010)\citenamefont {Zhang},
  \citenamefont {Be’er}, \citenamefont {Florin},\ and\ \citenamefont
  {Swinney}}]{zhang2010collective}%
  \BibitemOpen
  \bibfield  {author} {\bibinfo {author} {\bibfnamefont {H.-P.}\ \bibnamefont
  {Zhang}}, \bibinfo {author} {\bibfnamefont {A.}~\bibnamefont {Be’er}},
  \bibinfo {author} {\bibfnamefont {E.-L.}\ \bibnamefont {Florin}},\ and\
  \bibinfo {author} {\bibfnamefont {H.~L.}\ \bibnamefont {Swinney}},\
  }\bibfield  {title} {\bibinfo {title} {Collective motion and density
  fluctuations in bacterial colonies},\ }\href@noop {} {\bibfield  {journal}
  {\bibinfo  {journal} {Proceedings of the National Academy of Sciences}\
  }\textbf {\bibinfo {volume} {107}},\ \bibinfo {pages} {13626} (\bibinfo
  {year} {2010})}\BibitemShut {NoStop}%
\bibitem [{\citenamefont {Buttinoni}\ \emph {et~al.}(2013)\citenamefont
  {Buttinoni}, \citenamefont {Bialk{\'e}}, \citenamefont {K{\"u}mmel},
  \citenamefont {L{\"o}wen}, \citenamefont {Bechinger},\ and\ \citenamefont
  {Speck}}]{buttinoni2013dynamical}%
  \BibitemOpen
  \bibfield  {author} {\bibinfo {author} {\bibfnamefont {I.}~\bibnamefont
  {Buttinoni}}, \bibinfo {author} {\bibfnamefont {J.}~\bibnamefont
  {Bialk{\'e}}}, \bibinfo {author} {\bibfnamefont {F.}~\bibnamefont
  {K{\"u}mmel}}, \bibinfo {author} {\bibfnamefont {H.}~\bibnamefont
  {L{\"o}wen}}, \bibinfo {author} {\bibfnamefont {C.}~\bibnamefont
  {Bechinger}},\ and\ \bibinfo {author} {\bibfnamefont {T.}~\bibnamefont
  {Speck}},\ }\bibfield  {title} {\bibinfo {title} {Dynamical clustering and
  phase separation in suspensions of self-propelled colloidal particles},\
  }\href@noop {} {\bibfield  {journal} {\bibinfo  {journal} {Physical Review
  Letters}\ }\textbf {\bibinfo {volume} {110}},\ \bibinfo {pages} {238301}
  (\bibinfo {year} {2013})}\BibitemShut {NoStop}%
\bibitem [{\citenamefont {Cates}\ and\ \citenamefont
  {Tailleur}(2015)}]{cates2015motility}%
  \BibitemOpen
  \bibfield  {author} {\bibinfo {author} {\bibfnamefont {M.~E.}\ \bibnamefont
  {Cates}}\ and\ \bibinfo {author} {\bibfnamefont {J.}~\bibnamefont
  {Tailleur}},\ }\bibfield  {title} {\bibinfo {title} {Motility-induced phase
  separation},\ }\href@noop {} {\bibfield  {journal} {\bibinfo  {journal}
  {Annu. Rev. Condens. Matter Phys.}\ }\textbf {\bibinfo {volume} {6}},\
  \bibinfo {pages} {219} (\bibinfo {year} {2015})}\BibitemShut {NoStop}%
\bibitem [{\citenamefont {Zhang}\ \emph {et~al.}(2021)\citenamefont {Zhang},
  \citenamefont {Alert}, \citenamefont {Yan}, \citenamefont {Wingreen},\ and\
  \citenamefont {Granick}}]{zhang2021active}%
  \BibitemOpen
  \bibfield  {author} {\bibinfo {author} {\bibfnamefont {J.}~\bibnamefont
  {Zhang}}, \bibinfo {author} {\bibfnamefont {R.}~\bibnamefont {Alert}},
  \bibinfo {author} {\bibfnamefont {J.}~\bibnamefont {Yan}}, \bibinfo {author}
  {\bibfnamefont {N.~S.}\ \bibnamefont {Wingreen}},\ and\ \bibinfo {author}
  {\bibfnamefont {S.}~\bibnamefont {Granick}},\ }\bibfield  {title} {\bibinfo
  {title} {Active phase separation by turning towards regions of higher
  density},\ }\href@noop {} {\bibfield  {journal} {\bibinfo  {journal} {Nature
  Physics}\ }\textbf {\bibinfo {volume} {17}},\ \bibinfo {pages} {961}
  (\bibinfo {year} {2021})}\BibitemShut {NoStop}%
\bibitem [{\citenamefont {Chen}\ \emph {et~al.}(2023)\citenamefont {Chen},
  \citenamefont {Wang},\ and\ \citenamefont {Zhang}}]{chen2023tunable}%
  \BibitemOpen
  \bibfield  {author} {\bibinfo {author} {\bibfnamefont {Y.}~\bibnamefont
  {Chen}}, \bibinfo {author} {\bibfnamefont {L.}~\bibnamefont {Wang}},\ and\
  \bibinfo {author} {\bibfnamefont {T.~H.}\ \bibnamefont {Zhang}},\ }\bibfield
  {title} {\bibinfo {title} {Tunable collective dynamics of ellipsoidal quincke
  particles},\ }\href@noop {} {\bibfield  {journal} {\bibinfo  {journal} {Soft
  Matter}\ }\textbf {\bibinfo {volume} {19}},\ \bibinfo {pages} {512} (\bibinfo
  {year} {2023})}\BibitemShut {NoStop}%
\bibitem [{\citenamefont {Vogel}\ \emph {et~al.}(2015)\citenamefont {Vogel},
  \citenamefont {Utech}, \citenamefont {England}, \citenamefont {Shirman},
  \citenamefont {Phillips}, \citenamefont {Koay}, \citenamefont {Burgess},
  \citenamefont {Kolle}, \citenamefont {Weitz},\ and\ \citenamefont
  {Aizenberg}}]{vogel2015color}%
  \BibitemOpen
  \bibfield  {author} {\bibinfo {author} {\bibfnamefont {N.}~\bibnamefont
  {Vogel}}, \bibinfo {author} {\bibfnamefont {S.}~\bibnamefont {Utech}},
  \bibinfo {author} {\bibfnamefont {G.~T.}\ \bibnamefont {England}}, \bibinfo
  {author} {\bibfnamefont {T.}~\bibnamefont {Shirman}}, \bibinfo {author}
  {\bibfnamefont {K.~R.}\ \bibnamefont {Phillips}}, \bibinfo {author}
  {\bibfnamefont {N.}~\bibnamefont {Koay}}, \bibinfo {author} {\bibfnamefont
  {I.~B.}\ \bibnamefont {Burgess}}, \bibinfo {author} {\bibfnamefont
  {M.}~\bibnamefont {Kolle}}, \bibinfo {author} {\bibfnamefont {D.~A.}\
  \bibnamefont {Weitz}},\ and\ \bibinfo {author} {\bibfnamefont
  {J.}~\bibnamefont {Aizenberg}},\ }\bibfield  {title} {\bibinfo {title} {Color
  from hierarchy: Diverse optical properties of micron-sized spherical
  colloidal assemblies},\ }\href@noop {} {\bibfield  {journal} {\bibinfo
  {journal} {Proceedings of the National Academy of Sciences}\ }\textbf
  {\bibinfo {volume} {112}},\ \bibinfo {pages} {10845} (\bibinfo {year}
  {2015})}\BibitemShut {NoStop}%
\bibitem [{\citenamefont {Needleman}\ and\ \citenamefont
  {Dogic}(2017)}]{needleman2017active}%
  \BibitemOpen
  \bibfield  {author} {\bibinfo {author} {\bibfnamefont {D.}~\bibnamefont
  {Needleman}}\ and\ \bibinfo {author} {\bibfnamefont {Z.}~\bibnamefont
  {Dogic}},\ }\bibfield  {title} {\bibinfo {title} {Active matter at the
  interface between materials science and cell biology},\ }\href@noop {}
  {\bibfield  {journal} {\bibinfo  {journal} {Nature Reviews Materials}\
  }\textbf {\bibinfo {volume} {2}},\ \bibinfo {pages} {1} (\bibinfo {year}
  {2017})}\BibitemShut {NoStop}%
\bibitem [{\citenamefont {Geyer}\ \emph {et~al.}(2019)\citenamefont {Geyer},
  \citenamefont {Martin}, \citenamefont {Tailleur},\ and\ \citenamefont
  {Bartolo}}]{geyer2019freezing}%
  \BibitemOpen
  \bibfield  {author} {\bibinfo {author} {\bibfnamefont {D.}~\bibnamefont
  {Geyer}}, \bibinfo {author} {\bibfnamefont {D.}~\bibnamefont {Martin}},
  \bibinfo {author} {\bibfnamefont {J.}~\bibnamefont {Tailleur}},\ and\
  \bibinfo {author} {\bibfnamefont {D.}~\bibnamefont {Bartolo}},\ }\bibfield
  {title} {\bibinfo {title} {Freezing a flock: Motility-induced phase
  separation in polar active liquids},\ }\href@noop {} {\bibfield  {journal}
  {\bibinfo  {journal} {Physical Review X}\ }\textbf {\bibinfo {volume} {9}},\
  \bibinfo {pages} {031043} (\bibinfo {year} {2019})}\BibitemShut {NoStop}%
\bibitem [{\citenamefont {Digregorio}\ \emph {et~al.}(2018)\citenamefont
  {Digregorio}, \citenamefont {Levis}, \citenamefont {Suma}, \citenamefont
  {Cugliandolo}, \citenamefont {Gonnella},\ and\ \citenamefont
  {Pagonabarraga}}]{digregorio2018full}%
  \BibitemOpen
  \bibfield  {author} {\bibinfo {author} {\bibfnamefont {P.}~\bibnamefont
  {Digregorio}}, \bibinfo {author} {\bibfnamefont {D.}~\bibnamefont {Levis}},
  \bibinfo {author} {\bibfnamefont {A.}~\bibnamefont {Suma}}, \bibinfo {author}
  {\bibfnamefont {L.~F.}\ \bibnamefont {Cugliandolo}}, \bibinfo {author}
  {\bibfnamefont {G.}~\bibnamefont {Gonnella}},\ and\ \bibinfo {author}
  {\bibfnamefont {I.}~\bibnamefont {Pagonabarraga}},\ }\bibfield  {title}
  {\bibinfo {title} {Full phase diagram of active brownian disks: from melting
  to motility-induced phase separation},\ }\href@noop {} {\bibfield  {journal}
  {\bibinfo  {journal} {Physical Review Letters}\ }\textbf {\bibinfo {volume}
  {121}},\ \bibinfo {pages} {098003} (\bibinfo {year} {2018})}\BibitemShut
  {NoStop}%
\bibitem [{\citenamefont {Petroff}\ \emph {et~al.}(2015)\citenamefont
  {Petroff}, \citenamefont {Wu},\ and\ \citenamefont
  {Libchaber}}]{petroff2015fast}%
  \BibitemOpen
  \bibfield  {author} {\bibinfo {author} {\bibfnamefont {A.~P.}\ \bibnamefont
  {Petroff}}, \bibinfo {author} {\bibfnamefont {X.-L.}\ \bibnamefont {Wu}},\
  and\ \bibinfo {author} {\bibfnamefont {A.}~\bibnamefont {Libchaber}},\
  }\bibfield  {title} {\bibinfo {title} {Fast-moving bacteria self-organize
  into active two-dimensional crystals of rotating cells},\ }\href@noop {}
  {\bibfield  {journal} {\bibinfo  {journal} {Physical Review Letters}\
  }\textbf {\bibinfo {volume} {114}},\ \bibinfo {pages} {158102} (\bibinfo
  {year} {2015})}\BibitemShut {NoStop}%
\bibitem [{\citenamefont {Briand}\ and\ \citenamefont
  {Dauchot}(2016)}]{briand2016crystallization}%
  \BibitemOpen
  \bibfield  {author} {\bibinfo {author} {\bibfnamefont {G.}~\bibnamefont
  {Briand}}\ and\ \bibinfo {author} {\bibfnamefont {O.}~\bibnamefont
  {Dauchot}},\ }\bibfield  {title} {\bibinfo {title} {Crystallization of
  self-propelled hard discs},\ }\href@noop {} {\bibfield  {journal} {\bibinfo
  {journal} {Physical Review Letters}\ }\textbf {\bibinfo {volume} {117}},\
  \bibinfo {pages} {098004} (\bibinfo {year} {2016})}\BibitemShut {NoStop}%
\bibitem [{\citenamefont {Torquato}(2018)}]{torquato2018hyperuniform}%
  \BibitemOpen
  \bibfield  {author} {\bibinfo {author} {\bibfnamefont {S.}~\bibnamefont
  {Torquato}},\ }\bibfield  {title} {\bibinfo {title} {Hyperuniform states of
  matter},\ }\href@noop {} {\bibfield  {journal} {\bibinfo  {journal} {Physics
  Reports}\ }\textbf {\bibinfo {volume} {745}},\ \bibinfo {pages} {1} (\bibinfo
  {year} {2018})}\BibitemShut {NoStop}%
\bibitem [{\citenamefont {Yu}\ \emph {et~al.}(2021)\citenamefont {Yu},
  \citenamefont {Qiu}, \citenamefont {Chong}, \citenamefont {Torquato},\ and\
  \citenamefont {Park}}]{yu2021engineered}%
  \BibitemOpen
  \bibfield  {author} {\bibinfo {author} {\bibfnamefont {S.}~\bibnamefont
  {Yu}}, \bibinfo {author} {\bibfnamefont {C.-W.}\ \bibnamefont {Qiu}},
  \bibinfo {author} {\bibfnamefont {Y.}~\bibnamefont {Chong}}, \bibinfo
  {author} {\bibfnamefont {S.}~\bibnamefont {Torquato}},\ and\ \bibinfo
  {author} {\bibfnamefont {N.}~\bibnamefont {Park}},\ }\bibfield  {title}
  {\bibinfo {title} {Engineered disorder in photonics},\ }\href@noop {}
  {\bibfield  {journal} {\bibinfo  {journal} {Nature Reviews Materials}\
  }\textbf {\bibinfo {volume} {6}},\ \bibinfo {pages} {226} (\bibinfo {year}
  {2021})}\BibitemShut {NoStop}%
\bibitem [{\citenamefont {Lei}\ and\ \citenamefont {Ni}(2024)}]{lei2024non}%
  \BibitemOpen
  \bibfield  {author} {\bibinfo {author} {\bibfnamefont {Y.}~\bibnamefont
  {Lei}}\ and\ \bibinfo {author} {\bibfnamefont {R.}~\bibnamefont {Ni}},\
  }\bibfield  {title} {\bibinfo {title} {Non-equilibrium dynamic hyperuniform
  states},\ }\href@noop {} {\bibfield  {journal} {\bibinfo  {journal} {Journal
  of Physics: Condensed Matter}\ }\textbf {\bibinfo {volume} {37}},\ \bibinfo
  {pages} {023004} (\bibinfo {year} {2024})}\BibitemShut {NoStop}%
\bibitem [{\citenamefont {Lomba}\ \emph {et~al.}(2018)\citenamefont {Lomba},
  \citenamefont {Weis},\ and\ \citenamefont {Torquato}}]{lomba2018disordered}%
  \BibitemOpen
  \bibfield  {author} {\bibinfo {author} {\bibfnamefont {E.}~\bibnamefont
  {Lomba}}, \bibinfo {author} {\bibfnamefont {J.-J.}\ \bibnamefont {Weis}},\
  and\ \bibinfo {author} {\bibfnamefont {S.}~\bibnamefont {Torquato}},\
  }\bibfield  {title} {\bibinfo {title} {Disordered multihyperuniformity
  derived from binary plasmas},\ }\href@noop {} {\bibfield  {journal} {\bibinfo
   {journal} {Physical Review E}\ }\textbf {\bibinfo {volume} {97}},\ \bibinfo
  {pages} {010102} (\bibinfo {year} {2018})}\BibitemShut {NoStop}%
\bibitem [{\citenamefont {Lomba}\ \emph {et~al.}(2017)\citenamefont {Lomba},
  \citenamefont {Weis},\ and\ \citenamefont {Torquato}}]{lomba2017disordered}%
  \BibitemOpen
  \bibfield  {author} {\bibinfo {author} {\bibfnamefont {E.}~\bibnamefont
  {Lomba}}, \bibinfo {author} {\bibfnamefont {J.-J.}\ \bibnamefont {Weis}},\
  and\ \bibinfo {author} {\bibfnamefont {S.}~\bibnamefont {Torquato}},\
  }\bibfield  {title} {\bibinfo {title} {Disordered hyperuniformity in
  two-component nonadditive hard-disk plasmas},\ }\href@noop {} {\bibfield
  {journal} {\bibinfo  {journal} {Physical Review E}\ }\textbf {\bibinfo
  {volume} {96}},\ \bibinfo {pages} {062126} (\bibinfo {year}
  {2017})}\BibitemShut {NoStop}%
\bibitem [{\citenamefont {Zhang}\ and\ \citenamefont
  {Snezhko}(2022)}]{zhang2022hyperuniform}%
  \BibitemOpen
  \bibfield  {author} {\bibinfo {author} {\bibfnamefont {B.}~\bibnamefont
  {Zhang}}\ and\ \bibinfo {author} {\bibfnamefont {A.}~\bibnamefont
  {Snezhko}},\ }\bibfield  {title} {\bibinfo {title} {Hyperuniform active
  chiral fluids with tunable internal structure},\ }\href@noop {} {\bibfield
  {journal} {\bibinfo  {journal} {Physical Review Letters}\ }\textbf {\bibinfo
  {volume} {128}},\ \bibinfo {pages} {218002} (\bibinfo {year}
  {2022})}\BibitemShut {NoStop}%
\bibitem [{\citenamefont {Huang}\ \emph {et~al.}(2021)\citenamefont {Huang},
  \citenamefont {Hu}, \citenamefont {Yang}, \citenamefont {Liu},\ and\
  \citenamefont {Zhang}}]{huang2021circular}%
  \BibitemOpen
  \bibfield  {author} {\bibinfo {author} {\bibfnamefont {M.}~\bibnamefont
  {Huang}}, \bibinfo {author} {\bibfnamefont {W.}~\bibnamefont {Hu}}, \bibinfo
  {author} {\bibfnamefont {S.}~\bibnamefont {Yang}}, \bibinfo {author}
  {\bibfnamefont {Q.-X.}\ \bibnamefont {Liu}},\ and\ \bibinfo {author}
  {\bibfnamefont {H.}~\bibnamefont {Zhang}},\ }\bibfield  {title} {\bibinfo
  {title} {Circular swimming motility and disordered hyperuniform state in an
  algae system},\ }\href@noop {} {\bibfield  {journal} {\bibinfo  {journal}
  {Proceedings of the National Academy of Sciences}\ }\textbf {\bibinfo
  {volume} {118}},\ \bibinfo {pages} {e2100493118} (\bibinfo {year}
  {2021})}\BibitemShut {NoStop}%
\bibitem [{\citenamefont {Lei}\ \emph {et~al.}(2019)\citenamefont {Lei},
  \citenamefont {Ciamarra},\ and\ \citenamefont {Ni}}]{lei2019nonequilibrium}%
  \BibitemOpen
  \bibfield  {author} {\bibinfo {author} {\bibfnamefont {Q.-L.}\ \bibnamefont
  {Lei}}, \bibinfo {author} {\bibfnamefont {M.~P.}\ \bibnamefont {Ciamarra}},\
  and\ \bibinfo {author} {\bibfnamefont {R.}~\bibnamefont {Ni}},\ }\bibfield
  {title} {\bibinfo {title} {Nonequilibrium strongly hyperuniform fluids of
  circle active particles with large local density fluctuations},\ }\href@noop
  {} {\bibfield  {journal} {\bibinfo  {journal} {Science Advances}\ }\textbf
  {\bibinfo {volume} {5}},\ \bibinfo {pages} {eaau7423} (\bibinfo {year}
  {2019})}\BibitemShut {NoStop}%
\bibitem [{\citenamefont {Oppenheimer}\ \emph {et~al.}(2022)\citenamefont
  {Oppenheimer}, \citenamefont {Stein}, \citenamefont {Zion},\ and\
  \citenamefont {Shelley}}]{oppenheimer2022hyperuniformity}%
  \BibitemOpen
  \bibfield  {author} {\bibinfo {author} {\bibfnamefont {N.}~\bibnamefont
  {Oppenheimer}}, \bibinfo {author} {\bibfnamefont {D.~B.}\ \bibnamefont
  {Stein}}, \bibinfo {author} {\bibfnamefont {M.~Y.~B.}\ \bibnamefont {Zion}},\
  and\ \bibinfo {author} {\bibfnamefont {M.~J.}\ \bibnamefont {Shelley}},\
  }\bibfield  {title} {\bibinfo {title} {Hyperuniformity and phase enrichment
  in vortex and rotor assemblies},\ }\href@noop {} {\bibfield  {journal}
  {\bibinfo  {journal} {Nature Communications}\ }\textbf {\bibinfo {volume}
  {13}},\ \bibinfo {pages} {804} (\bibinfo {year} {2022})}\BibitemShut
  {NoStop}%
\bibitem [{\citenamefont {Corte}\ \emph {et~al.}(2008)\citenamefont {Corte},
  \citenamefont {Chaikin}, \citenamefont {Gollub},\ and\ \citenamefont
  {Pine}}]{corte2008random}%
  \BibitemOpen
  \bibfield  {author} {\bibinfo {author} {\bibfnamefont {L.}~\bibnamefont
  {Corte}}, \bibinfo {author} {\bibfnamefont {P.~M.}\ \bibnamefont {Chaikin}},
  \bibinfo {author} {\bibfnamefont {J.~P.}\ \bibnamefont {Gollub}},\ and\
  \bibinfo {author} {\bibfnamefont {D.~J.}\ \bibnamefont {Pine}},\ }\bibfield
  {title} {\bibinfo {title} {Random organization in periodically driven
  systems},\ }\href@noop {} {\bibfield  {journal} {\bibinfo  {journal} {Nature
  Physics}\ }\textbf {\bibinfo {volume} {4}},\ \bibinfo {pages} {420} (\bibinfo
  {year} {2008})}\BibitemShut {NoStop}%
\bibitem [{\citenamefont {Hexner}\ and\ \citenamefont
  {Levine}(2015)}]{hexner2015hyperuniformity}%
  \BibitemOpen
  \bibfield  {author} {\bibinfo {author} {\bibfnamefont {D.}~\bibnamefont
  {Hexner}}\ and\ \bibinfo {author} {\bibfnamefont {D.}~\bibnamefont
  {Levine}},\ }\bibfield  {title} {\bibinfo {title} {Hyperuniformity of
  critical absorbing states},\ }\href@noop {} {\bibfield  {journal} {\bibinfo
  {journal} {Physical Review Letters}\ }\textbf {\bibinfo {volume} {114}},\
  \bibinfo {pages} {110602} (\bibinfo {year} {2015})}\BibitemShut {NoStop}%
\bibitem [{\citenamefont {Tjhung}\ and\ \citenamefont
  {Berthier}(2015)}]{tjhung2015hyperuniform}%
  \BibitemOpen
  \bibfield  {author} {\bibinfo {author} {\bibfnamefont {E.}~\bibnamefont
  {Tjhung}}\ and\ \bibinfo {author} {\bibfnamefont {L.}~\bibnamefont
  {Berthier}},\ }\bibfield  {title} {\bibinfo {title} {Hyperuniform density
  fluctuations and diverging dynamic correlations in periodically driven
  colloidal suspensions},\ }\href@noop {} {\bibfield  {journal} {\bibinfo
  {journal} {Physical Review Letters}\ }\textbf {\bibinfo {volume} {114}},\
  \bibinfo {pages} {148301} (\bibinfo {year} {2015})}\BibitemShut {NoStop}%
\bibitem [{\citenamefont {Wang}\ \emph {et~al.}(2018)\citenamefont {Wang},
  \citenamefont {Schwarz},\ and\ \citenamefont
  {Paulsen}}]{wang2018hyperuniformity}%
  \BibitemOpen
  \bibfield  {author} {\bibinfo {author} {\bibfnamefont {J.}~\bibnamefont
  {Wang}}, \bibinfo {author} {\bibfnamefont {J.~M.}\ \bibnamefont {Schwarz}},\
  and\ \bibinfo {author} {\bibfnamefont {J.~D.}\ \bibnamefont {Paulsen}},\
  }\bibfield  {title} {\bibinfo {title} {Hyperuniformity with no fine tuning in
  sheared sedimenting suspensions},\ }\href@noop {} {\bibfield  {journal}
  {\bibinfo  {journal} {Nature Communications}\ }\textbf {\bibinfo {volume}
  {9}},\ \bibinfo {pages} {2836} (\bibinfo {year} {2018})}\BibitemShut
  {NoStop}%
\bibitem [{\citenamefont {Ishikawa}(2024)}]{ishikawa2024fluid}%
  \BibitemOpen
  \bibfield  {author} {\bibinfo {author} {\bibfnamefont {T.}~\bibnamefont
  {Ishikawa}},\ }\bibfield  {title} {\bibinfo {title} {Fluid dynamics of
  squirmers and ciliated microorganisms},\ }\href@noop {} {\bibfield  {journal}
  {\bibinfo  {journal} {Annual Review of Fluid Mechanics}\ }\textbf {\bibinfo
  {volume} {56}},\ \bibinfo {pages} {119} (\bibinfo {year} {2024})}\BibitemShut
  {NoStop}%
\bibitem [{\citenamefont {Moran}\ and\ \citenamefont
  {Posner}(2017)}]{moran2017phoretic}%
  \BibitemOpen
  \bibfield  {author} {\bibinfo {author} {\bibfnamefont {J.~L.}\ \bibnamefont
  {Moran}}\ and\ \bibinfo {author} {\bibfnamefont {J.~D.}\ \bibnamefont
  {Posner}},\ }\bibfield  {title} {\bibinfo {title} {Phoretic
  self-propulsion},\ }\href@noop {} {\bibfield  {journal} {\bibinfo  {journal}
  {Annual Review of Fluid Mechanics}\ }\textbf {\bibinfo {volume} {49}},\
  \bibinfo {pages} {511} (\bibinfo {year} {2017})}\BibitemShut {NoStop}%
\bibitem [{\citenamefont {Paxton}\ \emph {et~al.}(2006)\citenamefont {Paxton},
  \citenamefont {Baker}, \citenamefont {Kline}, \citenamefont {Wang},
  \citenamefont {Mallouk},\ and\ \citenamefont
  {Sen}}]{paxton2006catalytically}%
  \BibitemOpen
  \bibfield  {author} {\bibinfo {author} {\bibfnamefont {W.~F.}\ \bibnamefont
  {Paxton}}, \bibinfo {author} {\bibfnamefont {P.~T.}\ \bibnamefont {Baker}},
  \bibinfo {author} {\bibfnamefont {T.~R.}\ \bibnamefont {Kline}}, \bibinfo
  {author} {\bibfnamefont {Y.}~\bibnamefont {Wang}}, \bibinfo {author}
  {\bibfnamefont {T.~E.}\ \bibnamefont {Mallouk}},\ and\ \bibinfo {author}
  {\bibfnamefont {A.}~\bibnamefont {Sen}},\ }\bibfield  {title} {\bibinfo
  {title} {Catalytically induced electrokinetics for motors and micropumps},\
  }\href@noop {} {\bibfield  {journal} {\bibinfo  {journal} {Journal of the
  American Chemical Society}\ }\textbf {\bibinfo {volume} {128}},\ \bibinfo
  {pages} {14881} (\bibinfo {year} {2006})}\BibitemShut {NoStop}%
\bibitem [{\citenamefont {Kuron}\ \emph {et~al.}(2018)\citenamefont {Kuron},
  \citenamefont {Kreissl},\ and\ \citenamefont {Holm}}]{kuron2018toward}%
  \BibitemOpen
  \bibfield  {author} {\bibinfo {author} {\bibfnamefont {M.}~\bibnamefont
  {Kuron}}, \bibinfo {author} {\bibfnamefont {P.}~\bibnamefont {Kreissl}},\
  and\ \bibinfo {author} {\bibfnamefont {C.}~\bibnamefont {Holm}},\ }\bibfield
  {title} {\bibinfo {title} {Toward understanding of self-electrophoretic
  propulsion under realistic conditions: From bulk reactions to confinement
  effects},\ }\href@noop {} {\bibfield  {journal} {\bibinfo  {journal}
  {Accounts of Chemical Research}\ }\textbf {\bibinfo {volume} {51}},\ \bibinfo
  {pages} {2998} (\bibinfo {year} {2018})}\BibitemShut {NoStop}%
\bibitem [{\citenamefont {Jiang}\ \emph {et~al.}(2010)\citenamefont {Jiang},
  \citenamefont {Yoshinaga},\ and\ \citenamefont {Sano}}]{jiang2010active}%
  \BibitemOpen
  \bibfield  {author} {\bibinfo {author} {\bibfnamefont {H.-R.}\ \bibnamefont
  {Jiang}}, \bibinfo {author} {\bibfnamefont {N.}~\bibnamefont {Yoshinaga}},\
  and\ \bibinfo {author} {\bibfnamefont {M.}~\bibnamefont {Sano}},\ }\bibfield
  {title} {\bibinfo {title} {Active motion of a {J}anus particle by
  self-thermophoresis in a defocused laser beam},\ }\href@noop {} {\bibfield
  {journal} {\bibinfo  {journal} {Physical Review Letters}\ }\textbf {\bibinfo
  {volume} {105}},\ \bibinfo {pages} {268302} (\bibinfo {year}
  {2010})}\BibitemShut {NoStop}%
\bibitem [{\citenamefont {Howse}\ \emph {et~al.}(2007)\citenamefont {Howse},
  \citenamefont {Jones}, \citenamefont {Ryan}, \citenamefont {Gough},
  \citenamefont {Vafabakhsh},\ and\ \citenamefont
  {Golestanian}}]{howse2007self}%
  \BibitemOpen
  \bibfield  {author} {\bibinfo {author} {\bibfnamefont {J.~R.}\ \bibnamefont
  {Howse}}, \bibinfo {author} {\bibfnamefont {R.~A.}\ \bibnamefont {Jones}},
  \bibinfo {author} {\bibfnamefont {A.~J.}\ \bibnamefont {Ryan}}, \bibinfo
  {author} {\bibfnamefont {T.}~\bibnamefont {Gough}}, \bibinfo {author}
  {\bibfnamefont {R.}~\bibnamefont {Vafabakhsh}},\ and\ \bibinfo {author}
  {\bibfnamefont {R.}~\bibnamefont {Golestanian}},\ }\bibfield  {title}
  {\bibinfo {title} {Self-motile colloidal particles: from directed propulsion
  to random walk},\ }\href@noop {} {\bibfield  {journal} {\bibinfo  {journal}
  {Physical Review Letters}\ }\textbf {\bibinfo {volume} {99}},\ \bibinfo
  {pages} {048102} (\bibinfo {year} {2007})}\BibitemShut {NoStop}%
\bibitem [{\citenamefont {Anderson}(1989)}]{anderson1989colloid}%
  \BibitemOpen
  \bibfield  {author} {\bibinfo {author} {\bibfnamefont {J.~L.}\ \bibnamefont
  {Anderson}},\ }\bibfield  {title} {\bibinfo {title} {Colloid transport by
  interfacial forces},\ }\href@noop {} {\bibfield  {journal} {\bibinfo
  {journal} {Annual Review of Fluid Mechanics}\ }\textbf {\bibinfo {volume}
  {21}},\ \bibinfo {pages} {61} (\bibinfo {year} {1989})}\BibitemShut {NoStop}%
\bibitem [{\citenamefont {Lighthill}(1952)}]{lighthill1952squirming}%
  \BibitemOpen
  \bibfield  {author} {\bibinfo {author} {\bibfnamefont {M.~J.}\ \bibnamefont
  {Lighthill}},\ }\bibfield  {title} {\bibinfo {title} {On the squirming motion
  of nearly spherical deformable bodies through liquids at very small reynolds
  numbers},\ }\href@noop {} {\bibfield  {journal} {\bibinfo  {journal}
  {Communications on Pure and Applied mathematics}\ }\textbf {\bibinfo {volume}
  {5}},\ \bibinfo {pages} {109} (\bibinfo {year} {1952})}\BibitemShut {NoStop}%
\bibitem [{\citenamefont {Blake}(1971{\natexlab{a}})}]{Blake71}%
  \BibitemOpen
  \bibfield  {author} {\bibinfo {author} {\bibfnamefont {J.~R.}\ \bibnamefont
  {Blake}},\ }\bibfield  {title} {\bibinfo {title} {A spherical envelope
  approach to ciliary propulsion},\ }\href@noop {} {\bibfield  {journal}
  {\bibinfo  {journal} {Journal of Fluid Mechanics}\ }\textbf {\bibinfo
  {volume} {46}},\ \bibinfo {pages} {199} (\bibinfo {year}
  {1971}{\natexlab{a}})}\BibitemShut {NoStop}%
\bibitem [{\citenamefont {Popescu}\ \emph {et~al.}(2018)\citenamefont
  {Popescu}, \citenamefont {Uspal}, \citenamefont {Eskandari}, \citenamefont
  {Tasinkevych},\ and\ \citenamefont {Dietrich}}]{popescu2018effective}%
  \BibitemOpen
  \bibfield  {author} {\bibinfo {author} {\bibfnamefont {M.}~\bibnamefont
  {Popescu}}, \bibinfo {author} {\bibfnamefont {W.}~\bibnamefont {Uspal}},
  \bibinfo {author} {\bibfnamefont {Z.}~\bibnamefont {Eskandari}}, \bibinfo
  {author} {\bibfnamefont {M.}~\bibnamefont {Tasinkevych}},\ and\ \bibinfo
  {author} {\bibfnamefont {S.}~\bibnamefont {Dietrich}},\ }\bibfield  {title}
  {\bibinfo {title} {Effective squirmer models for self-phoretic chemically
  active spherical colloids},\ }\href@noop {} {\bibfield  {journal} {\bibinfo
  {journal} {The European Physical Journal E}\ }\textbf {\bibinfo {volume}
  {41}},\ \bibinfo {pages} {1} (\bibinfo {year} {2018})}\BibitemShut {NoStop}%
\bibitem [{\citenamefont {Poehnl}\ \emph {et~al.}(2020)\citenamefont {Poehnl},
  \citenamefont {Popescu},\ and\ \citenamefont
  {Uspal}}]{poehnl2020axisymmetric}%
  \BibitemOpen
  \bibfield  {author} {\bibinfo {author} {\bibfnamefont {R.}~\bibnamefont
  {Poehnl}}, \bibinfo {author} {\bibfnamefont {M.~N.}\ \bibnamefont
  {Popescu}},\ and\ \bibinfo {author} {\bibfnamefont {W.~E.}\ \bibnamefont
  {Uspal}},\ }\bibfield  {title} {\bibinfo {title} {Axisymmetric spheroidal
  squirmers and self-diffusiophoretic particles},\ }\href@noop {} {\bibfield
  {journal} {\bibinfo  {journal} {Journal of Physics: Condensed Matter}\
  }\textbf {\bibinfo {volume} {32}},\ \bibinfo {pages} {164001} (\bibinfo
  {year} {2020})}\BibitemShut {NoStop}%
\bibitem [{\citenamefont {Ishikawa}\ and\ \citenamefont
  {Pedley}(2008)}]{ishikawa2008coherent}%
  \BibitemOpen
  \bibfield  {author} {\bibinfo {author} {\bibfnamefont {T.}~\bibnamefont
  {Ishikawa}}\ and\ \bibinfo {author} {\bibfnamefont {T.~J.}\ \bibnamefont
  {Pedley}},\ }\bibfield  {title} {\bibinfo {title} {Coherent structures in
  monolayers of swimming particles},\ }\href@noop {} {\bibfield  {journal}
  {\bibinfo  {journal} {Physical Review Letters}\ }\textbf {\bibinfo {volume}
  {100}},\ \bibinfo {pages} {088103} (\bibinfo {year} {2008})}\BibitemShut
  {NoStop}%
\bibitem [{\citenamefont {Kuhr}\ \emph {et~al.}(2019)\citenamefont {Kuhr},
  \citenamefont {R{\"u}hle},\ and\ \citenamefont {Stark}}]{kuhr2019collective}%
  \BibitemOpen
  \bibfield  {author} {\bibinfo {author} {\bibfnamefont {J.-T.}\ \bibnamefont
  {Kuhr}}, \bibinfo {author} {\bibfnamefont {F.}~\bibnamefont {R{\"u}hle}},\
  and\ \bibinfo {author} {\bibfnamefont {H.}~\bibnamefont {Stark}},\ }\bibfield
   {title} {\bibinfo {title} {Collective dynamics in a monolayer of squirmers
  confined to a boundary by gravity},\ }\href@noop {} {\bibfield  {journal}
  {\bibinfo  {journal} {Soft Matter}\ }\textbf {\bibinfo {volume} {15}},\
  \bibinfo {pages} {5685} (\bibinfo {year} {2019})}\BibitemShut {NoStop}%
\bibitem [{\citenamefont {Kyoya}\ \emph {et~al.}(2015)\citenamefont {Kyoya},
  \citenamefont {Matsunaga}, \citenamefont {Imai}, \citenamefont {Omori},\ and\
  \citenamefont {Ishikawa}}]{kyoya2015shape}%
  \BibitemOpen
  \bibfield  {author} {\bibinfo {author} {\bibfnamefont {K.}~\bibnamefont
  {Kyoya}}, \bibinfo {author} {\bibfnamefont {D.}~\bibnamefont {Matsunaga}},
  \bibinfo {author} {\bibfnamefont {Y.}~\bibnamefont {Imai}}, \bibinfo {author}
  {\bibfnamefont {T.}~\bibnamefont {Omori}},\ and\ \bibinfo {author}
  {\bibfnamefont {T.}~\bibnamefont {Ishikawa}},\ }\bibfield  {title} {\bibinfo
  {title} {Shape matters: Near-field fluid mechanics dominate the collective
  motions of ellipsoidal squirmers},\ }\href@noop {} {\bibfield  {journal}
  {\bibinfo  {journal} {Physical Review E}\ }\textbf {\bibinfo {volume} {92}},\
  \bibinfo {pages} {063027} (\bibinfo {year} {2015})}\BibitemShut {NoStop}%
\bibitem [{\citenamefont {Zantop}\ and\ \citenamefont
  {Stark}(2022)}]{zantop2022emergent}%
  \BibitemOpen
  \bibfield  {author} {\bibinfo {author} {\bibfnamefont {A.~W.}\ \bibnamefont
  {Zantop}}\ and\ \bibinfo {author} {\bibfnamefont {H.}~\bibnamefont {Stark}},\
  }\bibfield  {title} {\bibinfo {title} {Emergent collective dynamics of pusher
  and puller squirmer rods: swarming, clustering, and turbulence},\ }\href@noop
  {} {\bibfield  {journal} {\bibinfo  {journal} {Soft Matter}\ }\textbf
  {\bibinfo {volume} {18}},\ \bibinfo {pages} {6179} (\bibinfo {year}
  {2022})}\BibitemShut {NoStop}%
\bibitem [{\citenamefont {Ishikawa}\ \emph {et~al.}(2006)\citenamefont
  {Ishikawa}, \citenamefont {Simmonds},\ and\ \citenamefont
  {Pedley}}]{ishikawa2006hydrodynamic}%
  \BibitemOpen
  \bibfield  {author} {\bibinfo {author} {\bibfnamefont {T.}~\bibnamefont
  {Ishikawa}}, \bibinfo {author} {\bibfnamefont {M.}~\bibnamefont {Simmonds}},\
  and\ \bibinfo {author} {\bibfnamefont {T.~J.}\ \bibnamefont {Pedley}},\
  }\bibfield  {title} {\bibinfo {title} {Hydrodynamic interaction of two
  swimming model micro-organisms},\ }\href@noop {} {\bibfield  {journal}
  {\bibinfo  {journal} {Journal of Fluid Mechanics}\ }\textbf {\bibinfo
  {volume} {568}},\ \bibinfo {pages} {119} (\bibinfo {year}
  {2006})}\BibitemShut {NoStop}%
\bibitem [{\citenamefont {Llopis}\ and\ \citenamefont
  {Pagonabarraga}(2010)}]{llopis2010hydrodynamic}%
  \BibitemOpen
  \bibfield  {author} {\bibinfo {author} {\bibfnamefont {I.}~\bibnamefont
  {Llopis}}\ and\ \bibinfo {author} {\bibfnamefont {I.}~\bibnamefont
  {Pagonabarraga}},\ }\bibfield  {title} {\bibinfo {title} {Hydrodynamic
  interactions in squirmer motion: Swimming with a neighbour and close to a
  wall},\ }\href@noop {} {\bibfield  {journal} {\bibinfo  {journal} {Journal of
  Non-Newtonian Fluid Mechanics}\ }\textbf {\bibinfo {volume} {165}},\ \bibinfo
  {pages} {946} (\bibinfo {year} {2010})}\BibitemShut {NoStop}%
\bibitem [{\citenamefont {Darveniza}\ \emph {et~al.}(2022)\citenamefont
  {Darveniza}, \citenamefont {Ishikawa}, \citenamefont {Pedley},\ and\
  \citenamefont {Brumley}}]{darveniza2022pairwise}%
  \BibitemOpen
  \bibfield  {author} {\bibinfo {author} {\bibfnamefont {C.}~\bibnamefont
  {Darveniza}}, \bibinfo {author} {\bibfnamefont {T.}~\bibnamefont {Ishikawa}},
  \bibinfo {author} {\bibfnamefont {T.}~\bibnamefont {Pedley}},\ and\ \bibinfo
  {author} {\bibfnamefont {D.}~\bibnamefont {Brumley}},\ }\bibfield  {title}
  {\bibinfo {title} {Pairwise scattering and bound states of spherical
  microorganisms},\ }\href@noop {} {\bibfield  {journal} {\bibinfo  {journal}
  {Physical Review Fluids}\ }\textbf {\bibinfo {volume} {7}},\ \bibinfo {pages}
  {013104} (\bibinfo {year} {2022})}\BibitemShut {NoStop}%
\bibitem [{\citenamefont {Theers}\ \emph {et~al.}(2016)\citenamefont {Theers},
  \citenamefont {Westphal}, \citenamefont {Gompper},\ and\ \citenamefont
  {Winkler}}]{theers2016modeling}%
  \BibitemOpen
  \bibfield  {author} {\bibinfo {author} {\bibfnamefont {M.}~\bibnamefont
  {Theers}}, \bibinfo {author} {\bibfnamefont {E.}~\bibnamefont {Westphal}},
  \bibinfo {author} {\bibfnamefont {G.}~\bibnamefont {Gompper}},\ and\ \bibinfo
  {author} {\bibfnamefont {R.~G.}\ \bibnamefont {Winkler}},\ }\bibfield
  {title} {\bibinfo {title} {Modeling a spheroidal microswimmer and cooperative
  swimming in a narrow slit},\ }\href@noop {} {\bibfield  {journal} {\bibinfo
  {journal} {Soft Matter}\ }\textbf {\bibinfo {volume} {12}},\ \bibinfo {pages}
  {7372} (\bibinfo {year} {2016})}\BibitemShut {NoStop}%
\bibitem [{\citenamefont {Pak}\ and\ \citenamefont
  {Lauga}(2014)}]{pak2014generalized}%
  \BibitemOpen
  \bibfield  {author} {\bibinfo {author} {\bibfnamefont {O.~S.}\ \bibnamefont
  {Pak}}\ and\ \bibinfo {author} {\bibfnamefont {E.}~\bibnamefont {Lauga}},\
  }\bibfield  {title} {\bibinfo {title} {Generalized squirming motion of a
  sphere},\ }\href@noop {} {\bibfield  {journal} {\bibinfo  {journal} {Journal
  of Engineering Mathematics}\ }\textbf {\bibinfo {volume} {88}},\ \bibinfo
  {pages} {1} (\bibinfo {year} {2014})}\BibitemShut {NoStop}%
\bibitem [{\citenamefont {Pedley}\ \emph {et~al.}(2016)\citenamefont {Pedley},
  \citenamefont {Brumley},\ and\ \citenamefont
  {Goldstein}}]{pedley2016squirmers}%
  \BibitemOpen
  \bibfield  {author} {\bibinfo {author} {\bibfnamefont {T.~J.}\ \bibnamefont
  {Pedley}}, \bibinfo {author} {\bibfnamefont {D.~R.}\ \bibnamefont
  {Brumley}},\ and\ \bibinfo {author} {\bibfnamefont {R.~E.}\ \bibnamefont
  {Goldstein}},\ }\bibfield  {title} {\bibinfo {title} {Squirmers with swirl: a
  model for volvox swimming},\ }\href@noop {} {\bibfield  {journal} {\bibinfo
  {journal} {Journal of Fluid Mechanics}\ }\textbf {\bibinfo {volume} {798}},\
  \bibinfo {pages} {165} (\bibinfo {year} {2016})}\BibitemShut {NoStop}%
\bibitem [{\citenamefont {Burada}\ \emph {et~al.}(2022)\citenamefont {Burada},
  \citenamefont {Maity},\ and\ \citenamefont
  {J{\"u}licher}}]{burada2022hydrodynamics}%
  \BibitemOpen
  \bibfield  {author} {\bibinfo {author} {\bibfnamefont {P.}~\bibnamefont
  {Burada}}, \bibinfo {author} {\bibfnamefont {R.}~\bibnamefont {Maity}},\ and\
  \bibinfo {author} {\bibfnamefont {F.}~\bibnamefont {J{\"u}licher}},\
  }\bibfield  {title} {\bibinfo {title} {Hydrodynamics of chiral squirmers},\
  }\href@noop {} {\bibfield  {journal} {\bibinfo  {journal} {Physical Review
  E}\ }\textbf {\bibinfo {volume} {105}},\ \bibinfo {pages} {024603} (\bibinfo
  {year} {2022})}\BibitemShut {NoStop}%
\bibitem [{\citenamefont {Samatas}\ and\ \citenamefont
  {Lintuvuori}(2023)}]{samatas2023hydrodynamic}%
  \BibitemOpen
  \bibfield  {author} {\bibinfo {author} {\bibfnamefont {S.}~\bibnamefont
  {Samatas}}\ and\ \bibinfo {author} {\bibfnamefont {J.}~\bibnamefont
  {Lintuvuori}},\ }\bibfield  {title} {\bibinfo {title} {Hydrodynamic
  synchronization of chiral microswimmers},\ }\href@noop {} {\bibfield
  {journal} {\bibinfo  {journal} {Physical Review Letters}\ }\textbf {\bibinfo
  {volume} {130}},\ \bibinfo {pages} {024001} (\bibinfo {year}
  {2023})}\BibitemShut {NoStop}%
\bibitem [{\citenamefont {Poehnl}\ and\ \citenamefont
  {Uspal}(2023)}]{poehnl2023shape}%
  \BibitemOpen
  \bibfield  {author} {\bibinfo {author} {\bibfnamefont {R.}~\bibnamefont
  {Poehnl}}\ and\ \bibinfo {author} {\bibfnamefont {W.~E.}\ \bibnamefont
  {Uspal}},\ }\bibfield  {title} {\bibinfo {title} {Shape-induced pairing of
  spheroidal squirmers},\ }\href@noop {} {\bibfield  {journal} {\bibinfo
  {journal} {Physical Review Fluids}\ }\textbf {\bibinfo {volume} {8}},\
  \bibinfo {pages} {113103} (\bibinfo {year} {2023})}\BibitemShut {NoStop}%
\bibitem [{\citenamefont {Katuri}\ \emph {et~al.}(2022)\citenamefont {Katuri},
  \citenamefont {Poehnl}, \citenamefont {Sokolov}, \citenamefont {Uspal},\ and\
  \citenamefont {Snezhko}}]{katuri2022arrested}%
  \BibitemOpen
  \bibfield  {author} {\bibinfo {author} {\bibfnamefont {J.}~\bibnamefont
  {Katuri}}, \bibinfo {author} {\bibfnamefont {R.}~\bibnamefont {Poehnl}},
  \bibinfo {author} {\bibfnamefont {A.}~\bibnamefont {Sokolov}}, \bibinfo
  {author} {\bibfnamefont {W.}~\bibnamefont {Uspal}},\ and\ \bibinfo {author}
  {\bibfnamefont {A.}~\bibnamefont {Snezhko}},\ }\bibfield  {title} {\bibinfo
  {title} {Arrested-motility states in populations of shape-anisotropic active
  {J}anus particles},\ }\href@noop {} {\bibfield  {journal} {\bibinfo
  {journal} {Science Advances}\ }\textbf {\bibinfo {volume} {8}},\ \bibinfo
  {pages} {eabo3604} (\bibinfo {year} {2022})}\BibitemShut {NoStop}%
\bibitem [{\citenamefont {Squires}\ and\ \citenamefont
  {Bazant}(2006)}]{squires2006breaking}%
  \BibitemOpen
  \bibfield  {author} {\bibinfo {author} {\bibfnamefont {T.~M.}\ \bibnamefont
  {Squires}}\ and\ \bibinfo {author} {\bibfnamefont {M.~Z.}\ \bibnamefont
  {Bazant}},\ }\bibfield  {title} {\bibinfo {title} {Breaking symmetries in
  induced-charge electro-osmosis and electrophoresis},\ }\href@noop {}
  {\bibfield  {journal} {\bibinfo  {journal} {Journal of Fluid Mechanics}\
  }\textbf {\bibinfo {volume} {560}},\ \bibinfo {pages} {65} (\bibinfo {year}
  {2006})}\BibitemShut {NoStop}%
\bibitem [{\citenamefont {Archer}\ \emph {et~al.}(2015)\citenamefont {Archer},
  \citenamefont {Campbell},\ and\ \citenamefont {Ebbens}}]{archer2015glancing}%
  \BibitemOpen
  \bibfield  {author} {\bibinfo {author} {\bibfnamefont {R.}~\bibnamefont
  {Archer}}, \bibinfo {author} {\bibfnamefont {A.}~\bibnamefont {Campbell}},\
  and\ \bibinfo {author} {\bibfnamefont {S.}~\bibnamefont {Ebbens}},\
  }\bibfield  {title} {\bibinfo {title} {Glancing angle metal evaporation
  synthesis of catalytic swimming {J}anus colloids with well defined angular
  velocity},\ }\href@noop {} {\bibfield  {journal} {\bibinfo  {journal} {Soft
  Matter}\ }\textbf {\bibinfo {volume} {11}},\ \bibinfo {pages} {6872}
  (\bibinfo {year} {2015})}\BibitemShut {NoStop}%
\bibitem [{\citenamefont {Uspal}\ \emph {et~al.}(2015)\citenamefont {Uspal},
  \citenamefont {Popescu}, \citenamefont {Dietrich},\ and\ \citenamefont
  {Tasinkevych}}]{uspal2015self}%
  \BibitemOpen
  \bibfield  {author} {\bibinfo {author} {\bibfnamefont {W.}~\bibnamefont
  {Uspal}}, \bibinfo {author} {\bibfnamefont {M.~N.}\ \bibnamefont {Popescu}},
  \bibinfo {author} {\bibfnamefont {S.}~\bibnamefont {Dietrich}},\ and\
  \bibinfo {author} {\bibfnamefont {M.}~\bibnamefont {Tasinkevych}},\
  }\bibfield  {title} {\bibinfo {title} {Self-propulsion of a catalytically
  active particle near a planar wall: from reflection to sliding and
  hovering},\ }\href@noop {} {\bibfield  {journal} {\bibinfo  {journal} {Soft
  Matter}\ }\textbf {\bibinfo {volume} {11}},\ \bibinfo {pages} {434} (\bibinfo
  {year} {2015})}\BibitemShut {NoStop}%
\bibitem [{\citenamefont {Florescu}\ \emph {et~al.}(2009)\citenamefont
  {Florescu}, \citenamefont {Torquato},\ and\ \citenamefont
  {Steinhardt}}]{florescu2009complete}%
  \BibitemOpen
  \bibfield  {author} {\bibinfo {author} {\bibfnamefont {M.}~\bibnamefont
  {Florescu}}, \bibinfo {author} {\bibfnamefont {S.}~\bibnamefont {Torquato}},\
  and\ \bibinfo {author} {\bibfnamefont {P.~J.}\ \bibnamefont {Steinhardt}},\
  }\bibfield  {title} {\bibinfo {title} {Complete band gaps in two-dimensional
  photonic quasicrystals},\ }\href@noop {} {\bibfield  {journal} {\bibinfo
  {journal} {Physical Review B—Condensed Matter and Materials Physics}\
  }\textbf {\bibinfo {volume} {80}},\ \bibinfo {pages} {155112} (\bibinfo
  {year} {2009})}\BibitemShut {NoStop}%
\bibitem [{\citenamefont {Froufe-P{\'e}rez}\ \emph {et~al.}(2017)\citenamefont
  {Froufe-P{\'e}rez}, \citenamefont {Engel}, \citenamefont {S{\'a}enz},\ and\
  \citenamefont {Scheffold}}]{froufe2017band}%
  \BibitemOpen
  \bibfield  {author} {\bibinfo {author} {\bibfnamefont {L.~S.}\ \bibnamefont
  {Froufe-P{\'e}rez}}, \bibinfo {author} {\bibfnamefont {M.}~\bibnamefont
  {Engel}}, \bibinfo {author} {\bibfnamefont {J.~J.}\ \bibnamefont
  {S{\'a}enz}},\ and\ \bibinfo {author} {\bibfnamefont {F.}~\bibnamefont
  {Scheffold}},\ }\bibfield  {title} {\bibinfo {title} {Band gap formation and
  anderson localization in disordered photonic materials with structural
  correlations},\ }\href@noop {} {\bibfield  {journal} {\bibinfo  {journal}
  {Proceedings of the National Academy of Sciences}\ }\textbf {\bibinfo
  {volume} {114}},\ \bibinfo {pages} {9570} (\bibinfo {year}
  {2017})}\BibitemShut {NoStop}%
\bibitem [{\citenamefont {Zhou}\ \emph {et~al.}(2019)\citenamefont {Zhou},
  \citenamefont {Tong}, \citenamefont {Sun},\ and\ \citenamefont
  {Tsang}}]{zhou2019hyperuniform}%
  \BibitemOpen
  \bibfield  {author} {\bibinfo {author} {\bibfnamefont {W.}~\bibnamefont
  {Zhou}}, \bibinfo {author} {\bibfnamefont {Y.}~\bibnamefont {Tong}}, \bibinfo
  {author} {\bibfnamefont {X.}~\bibnamefont {Sun}},\ and\ \bibinfo {author}
  {\bibfnamefont {H.~K.}\ \bibnamefont {Tsang}},\ }\bibfield  {title} {\bibinfo
  {title} {Hyperuniform disordered photonic bandgap polarizers},\ }\href@noop
  {} {\bibfield  {journal} {\bibinfo  {journal} {Journal of Applied Physics}\
  }\textbf {\bibinfo {volume} {126}} (\bibinfo {year} {2019})}\BibitemShut
  {NoStop}%
\bibitem [{\citenamefont {Ch{\'e}ron}\ \emph {et~al.}(2022)\citenamefont
  {Ch{\'e}ron}, \citenamefont {Groby}, \citenamefont {Pagneux}, \citenamefont
  {F{\'e}lix},\ and\ \citenamefont
  {Romero-Garc{\'\i}a}}]{cheron2022experimental}%
  \BibitemOpen
  \bibfield  {author} {\bibinfo {author} {\bibfnamefont {E.}~\bibnamefont
  {Ch{\'e}ron}}, \bibinfo {author} {\bibfnamefont {J.-P.}\ \bibnamefont
  {Groby}}, \bibinfo {author} {\bibfnamefont {V.}~\bibnamefont {Pagneux}},
  \bibinfo {author} {\bibfnamefont {S.}~\bibnamefont {F{\'e}lix}},\ and\
  \bibinfo {author} {\bibfnamefont {V.}~\bibnamefont {Romero-Garc{\'\i}a}},\
  }\bibfield  {title} {\bibinfo {title} {Experimental characterization of
  rigid-scatterer hyperuniform distributions for audible acoustics},\
  }\href@noop {} {\bibfield  {journal} {\bibinfo  {journal} {Physical Review
  B}\ }\textbf {\bibinfo {volume} {106}},\ \bibinfo {pages} {064206} (\bibinfo
  {year} {2022})}\BibitemShut {NoStop}%
\bibitem [{\citenamefont {Saintillan}\ and\ \citenamefont
  {Shelley}(2007)}]{saintillan2007orientational}%
  \BibitemOpen
  \bibfield  {author} {\bibinfo {author} {\bibfnamefont {D.}~\bibnamefont
  {Saintillan}}\ and\ \bibinfo {author} {\bibfnamefont {M.~J.}\ \bibnamefont
  {Shelley}},\ }\bibfield  {title} {\bibinfo {title} {Orientational order and
  instabilities in suspensions of self-locomoting rods},\ }\href@noop {}
  {\bibfield  {journal} {\bibinfo  {journal} {Physical Review Letters}\
  }\textbf {\bibinfo {volume} {99}},\ \bibinfo {pages} {058102} (\bibinfo
  {year} {2007})}\BibitemShut {NoStop}%
\bibitem [{\citenamefont {Saintillan}\ and\ \citenamefont
  {Shelley}(2008)}]{saintillan2008instabilities}%
  \BibitemOpen
  \bibfield  {author} {\bibinfo {author} {\bibfnamefont {D.}~\bibnamefont
  {Saintillan}}\ and\ \bibinfo {author} {\bibfnamefont {M.~J.}\ \bibnamefont
  {Shelley}},\ }\bibfield  {title} {\bibinfo {title} {Instabilities, pattern
  formation, and mixing in active suspensions},\ }\href@noop {} {\bibfield
  {journal} {\bibinfo  {journal} {Physics of Fluids}\ }\textbf {\bibinfo
  {volume} {20}} (\bibinfo {year} {2008})}\BibitemShut {NoStop}%
\bibitem [{\citenamefont {Saintillan}(2018)}]{saintillan2018rheology}%
  \BibitemOpen
  \bibfield  {author} {\bibinfo {author} {\bibfnamefont {D.}~\bibnamefont
  {Saintillan}},\ }\bibfield  {title} {\bibinfo {title} {Rheology of active
  fluids},\ }\href@noop {} {\bibfield  {journal} {\bibinfo  {journal} {Annual
  Review of Fluid Mechanics}\ }\textbf {\bibinfo {volume} {50}},\ \bibinfo
  {pages} {563} (\bibinfo {year} {2018})}\BibitemShut {NoStop}%
\bibitem [{\citenamefont {Graham}(2018)}]{graham2018microhydrodynamics}%
  \BibitemOpen
  \bibfield  {author} {\bibinfo {author} {\bibfnamefont {M.~D.}\ \bibnamefont
  {Graham}},\ }\href@noop {} {\emph {\bibinfo {title} {Microhydrodynamics,
  Brownian motion, and complex fluids}}},\ Vol.~\bibinfo {volume} {58}\
  (\bibinfo  {publisher} {Cambridge University Press},\ \bibinfo {year}
  {2018})\BibitemShut {NoStop}%
\bibitem [{\citenamefont {Heyes}\ and\ \citenamefont
  {Melrose}(1993)}]{heyes1993brownian}%
  \BibitemOpen
  \bibfield  {author} {\bibinfo {author} {\bibfnamefont {D.}~\bibnamefont
  {Heyes}}\ and\ \bibinfo {author} {\bibfnamefont {J.}~\bibnamefont
  {Melrose}},\ }\bibfield  {title} {\bibinfo {title} {Brownian dynamics
  simulations of model hard-sphere suspensions},\ }\href@noop {} {\bibfield
  {journal} {\bibinfo  {journal} {Journal of Non-Newtonian Fluid Mechanics}\
  }\textbf {\bibinfo {volume} {46}},\ \bibinfo {pages} {1} (\bibinfo {year}
  {1993})}\BibitemShut {NoStop}%
\bibitem [{\citenamefont {Frenkel}\ and\ \citenamefont
  {Smit}(2002)}]{frenkel2002understanding}%
  \BibitemOpen
  \bibfield  {author} {\bibinfo {author} {\bibfnamefont {D.}~\bibnamefont
  {Frenkel}}\ and\ \bibinfo {author} {\bibfnamefont {B.}~\bibnamefont {Smit}},\
  }\href@noop {} {\emph {\bibinfo {title} {Understanding molecular
  simulation}}}\ (\bibinfo  {publisher} {Academic Press},\ \bibinfo {address}
  {San Diego},\ \bibinfo {year} {2002})\BibitemShut {NoStop}%
\bibitem [{\citenamefont {Pozrikidis}(2002)}]{pozrikidis02}%
  \BibitemOpen
  \bibfield  {author} {\bibinfo {author} {\bibfnamefont {C.}~\bibnamefont
  {Pozrikidis}},\ }\href@noop {} {\emph {\bibinfo {title} {A Practical Guide to
  Boundary Element Methods with the Software Library BEMLIB}}}\ (\bibinfo
  {publisher} {CRC Press},\ \bibinfo {address} {Boca Raton},\ \bibinfo {year}
  {2002})\BibitemShut {NoStop}%
\bibitem [{\citenamefont {Hawat}\ \emph {et~al.}(2023)\citenamefont {Hawat},
  \citenamefont {Gautier}, \citenamefont {Bardenet},\ and\ \citenamefont
  {Lachi{\`e}ze-Rey}}]{hawat2023estimating}%
  \BibitemOpen
  \bibfield  {author} {\bibinfo {author} {\bibfnamefont {D.}~\bibnamefont
  {Hawat}}, \bibinfo {author} {\bibfnamefont {G.}~\bibnamefont {Gautier}},
  \bibinfo {author} {\bibfnamefont {R.}~\bibnamefont {Bardenet}},\ and\
  \bibinfo {author} {\bibfnamefont {R.}~\bibnamefont {Lachi{\`e}ze-Rey}},\
  }\bibfield  {title} {\bibinfo {title} {On estimating the structure factor of
  a point process, with applications to hyperuniformity},\ }\href@noop {}
  {\bibfield  {journal} {\bibinfo  {journal} {Statistics and Computing}\
  }\textbf {\bibinfo {volume} {33}},\ \bibinfo {pages} {61} (\bibinfo {year}
  {2023})}\BibitemShut {NoStop}%
\bibitem [{\citenamefont {Philcox}\ and\ \citenamefont
  {Torquato}(2023)}]{philcox2023disordered}%
  \BibitemOpen
  \bibfield  {author} {\bibinfo {author} {\bibfnamefont {O.~H.}\ \bibnamefont
  {Philcox}}\ and\ \bibinfo {author} {\bibfnamefont {S.}~\bibnamefont
  {Torquato}},\ }\bibfield  {title} {\bibinfo {title} {Disordered heterogeneous
  universe: Galaxy distribution and clustering across length scales},\
  }\href@noop {} {\bibfield  {journal} {\bibinfo  {journal} {Physical Review
  X}\ }\textbf {\bibinfo {volume} {13}},\ \bibinfo {pages} {011038} (\bibinfo
  {year} {2023})}\BibitemShut {NoStop}%
\bibitem [{\citenamefont {Press}\ \emph {et~al.}(1992)\citenamefont {Press},
  \citenamefont {Teukolsky}, \citenamefont {Vetterling},\ and\ \citenamefont
  {Flannery}}]{press1992numerical}%
  \BibitemOpen
  \bibfield  {author} {\bibinfo {author} {\bibfnamefont {W.~H.}\ \bibnamefont
  {Press}}, \bibinfo {author} {\bibfnamefont {S.~A.}\ \bibnamefont
  {Teukolsky}}, \bibinfo {author} {\bibfnamefont {W.~T.}\ \bibnamefont
  {Vetterling}},\ and\ \bibinfo {author} {\bibfnamefont {B.~P.}\ \bibnamefont
  {Flannery}},\ }\href@noop {} {\emph {\bibinfo {title} {Numerical Recipes}}}\
  (\bibinfo  {publisher} {Cambridge Univ. Press},\ \bibinfo {year}
  {1992})\BibitemShut {NoStop}%
\bibitem [{\citenamefont {Morris}\ \emph {et~al.}(1996)\citenamefont {Morris},
  \citenamefont {Bodenschatz}, \citenamefont {Cannell},\ and\ \citenamefont
  {Ahlers}}]{morris1996spatio}%
  \BibitemOpen
  \bibfield  {author} {\bibinfo {author} {\bibfnamefont {S.~W.}\ \bibnamefont
  {Morris}}, \bibinfo {author} {\bibfnamefont {E.}~\bibnamefont {Bodenschatz}},
  \bibinfo {author} {\bibfnamefont {D.~S.}\ \bibnamefont {Cannell}},\ and\
  \bibinfo {author} {\bibfnamefont {G.}~\bibnamefont {Ahlers}},\ }\bibfield
  {title} {\bibinfo {title} {The spatio-temporal structure of spiral-defect
  chaos},\ }\href@noop {} {\bibfield  {journal} {\bibinfo  {journal} {Physica
  D: Nonlinear Phenomena}\ }\textbf {\bibinfo {volume} {97}},\ \bibinfo {pages}
  {164} (\bibinfo {year} {1996})}\BibitemShut {NoStop}%
\bibitem [{\citenamefont {Cheng}\ \emph {et~al.}(2003)\citenamefont {Cheng},
  \citenamefont {Fenter}, \citenamefont {Bedzyk},\ and\ \citenamefont
  {Sturchio}}]{cheng2003fourier}%
  \BibitemOpen
  \bibfield  {author} {\bibinfo {author} {\bibfnamefont {L.}~\bibnamefont
  {Cheng}}, \bibinfo {author} {\bibfnamefont {P.}~\bibnamefont {Fenter}},
  \bibinfo {author} {\bibfnamefont {M.}~\bibnamefont {Bedzyk}},\ and\ \bibinfo
  {author} {\bibfnamefont {N.}~\bibnamefont {Sturchio}},\ }\bibfield  {title}
  {\bibinfo {title} {Fourier-expansion solution of atom distributions in a
  crystal using x-ray standing waves},\ }\href@noop {} {\bibfield  {journal}
  {\bibinfo  {journal} {Physical Review Letters}\ }\textbf {\bibinfo {volume}
  {90}},\ \bibinfo {pages} {255503} (\bibinfo {year} {2003})}\BibitemShut
  {NoStop}%
\bibitem [{\citenamefont {Soper}\ and\ \citenamefont
  {Barney}(2012)}]{soper2012use}%
  \BibitemOpen
  \bibfield  {author} {\bibinfo {author} {\bibfnamefont {A.~K.}\ \bibnamefont
  {Soper}}\ and\ \bibinfo {author} {\bibfnamefont {E.~R.}\ \bibnamefont
  {Barney}},\ }\bibfield  {title} {\bibinfo {title} {On the use of modification
  functions when fourier transforming total scattering data},\ }\href@noop {}
  {\bibfield  {journal} {\bibinfo  {journal} {Applied Crystallography}\
  }\textbf {\bibinfo {volume} {45}},\ \bibinfo {pages} {1314} (\bibinfo {year}
  {2012})}\BibitemShut {NoStop}%
\bibitem [{\citenamefont {Morse}\ \emph {et~al.}(2023)\citenamefont {Morse},
  \citenamefont {Kim}, \citenamefont {Steinhardt},\ and\ \citenamefont
  {Torquato}}]{morse2023generating}%
  \BibitemOpen
  \bibfield  {author} {\bibinfo {author} {\bibfnamefont {P.~K.}\ \bibnamefont
  {Morse}}, \bibinfo {author} {\bibfnamefont {J.}~\bibnamefont {Kim}}, \bibinfo
  {author} {\bibfnamefont {P.~J.}\ \bibnamefont {Steinhardt}},\ and\ \bibinfo
  {author} {\bibfnamefont {S.}~\bibnamefont {Torquato}},\ }\bibfield  {title}
  {\bibinfo {title} {Generating large disordered stealthy hyperuniform systems
  with ultrahigh accuracy to determine their physical properties},\ }\href@noop
  {} {\bibfield  {journal} {\bibinfo  {journal} {Physical Review Research}\
  }\textbf {\bibinfo {volume} {5}},\ \bibinfo {pages} {033190} (\bibinfo {year}
  {2023})}\BibitemShut {NoStop}%
\bibitem [{\citenamefont {Mecke}\ \emph {et~al.}(2024)\citenamefont {Mecke},
  \citenamefont {Nketsiah}, \citenamefont {Li},\ and\ \citenamefont
  {Gao}}]{mecke2024emergent}%
  \BibitemOpen
  \bibfield  {author} {\bibinfo {author} {\bibfnamefont {J.}~\bibnamefont
  {Mecke}}, \bibinfo {author} {\bibfnamefont {J.~O.}\ \bibnamefont {Nketsiah}},
  \bibinfo {author} {\bibfnamefont {R.}~\bibnamefont {Li}},\ and\ \bibinfo
  {author} {\bibfnamefont {Y.}~\bibnamefont {Gao}},\ }\bibfield  {title}
  {\bibinfo {title} {Emergent phenomena in chiral active matter},\ }\href@noop
  {} {\bibfield  {journal} {\bibinfo  {journal} {National Science Open}\
  }\textbf {\bibinfo {volume} {3}},\ \bibinfo {pages} {20230086} (\bibinfo
  {year} {2024})}\BibitemShut {NoStop}%
\bibitem [{\citenamefont {Imperio}\ and\ \citenamefont
  {Reatto}(2004)}]{imperio2004bidimensional}%
  \BibitemOpen
  \bibfield  {author} {\bibinfo {author} {\bibfnamefont {A.}~\bibnamefont
  {Imperio}}\ and\ \bibinfo {author} {\bibfnamefont {L.}~\bibnamefont
  {Reatto}},\ }\bibfield  {title} {\bibinfo {title} {A bidimensional fluid
  system with competing interactions: spontaneous and inducedpattern
  formation},\ }\href@noop {} {\bibfield  {journal} {\bibinfo  {journal}
  {Journal of Physics: Condensed matter}\ }\textbf {\bibinfo {volume} {16}},\
  \bibinfo {pages} {S3769} (\bibinfo {year} {2004})}\BibitemShut {NoStop}%
\bibitem [{\citenamefont {D{\'\i}az-Pozuelo}\ \emph {et~al.}(2025)\citenamefont
  {D{\'\i}az-Pozuelo}, \citenamefont {Gonz{\'a}lez-Salgado},\ and\
  \citenamefont {Lomba}}]{diaz2025build}%
  \BibitemOpen
  \bibfield  {author} {\bibinfo {author} {\bibfnamefont {A.}~\bibnamefont
  {D{\'\i}az-Pozuelo}}, \bibinfo {author} {\bibfnamefont {D.}~\bibnamefont
  {Gonz{\'a}lez-Salgado}},\ and\ \bibinfo {author} {\bibfnamefont
  {E.}~\bibnamefont {Lomba}},\ }\bibfield  {title} {\bibinfo {title} {On the
  build-up of effective hyperuniformity from large globular colloidal
  aggregates},\ }\href@noop {} {\bibfield  {journal} {\bibinfo  {journal} {The
  Journal of Chemical Physics}\ }\textbf {\bibinfo {volume} {162}} (\bibinfo
  {year} {2025})}\BibitemShut {NoStop}%
\bibitem [{\citenamefont {Alarc{\'o}n}\ and\ \citenamefont
  {Pagonabarraga}(2013)}]{alarcon2013spontaneous}%
  \BibitemOpen
  \bibfield  {author} {\bibinfo {author} {\bibfnamefont {F.}~\bibnamefont
  {Alarc{\'o}n}}\ and\ \bibinfo {author} {\bibfnamefont {I.}~\bibnamefont
  {Pagonabarraga}},\ }\bibfield  {title} {\bibinfo {title} {Spontaneous
  aggregation and global polar ordering in squirmer suspensions},\ }\href@noop
  {} {\bibfield  {journal} {\bibinfo  {journal} {Journal of Molecular Liquids}\
  }\textbf {\bibinfo {volume} {185}},\ \bibinfo {pages} {56} (\bibinfo {year}
  {2013})}\BibitemShut {NoStop}%
\bibitem [{\citenamefont {Oyama}\ \emph {et~al.}(2016)\citenamefont {Oyama},
  \citenamefont {Molina},\ and\ \citenamefont {Yamamoto}}]{oyama2016purely}%
  \BibitemOpen
  \bibfield  {author} {\bibinfo {author} {\bibfnamefont {N.}~\bibnamefont
  {Oyama}}, \bibinfo {author} {\bibfnamefont {J.~J.}\ \bibnamefont {Molina}},\
  and\ \bibinfo {author} {\bibfnamefont {R.}~\bibnamefont {Yamamoto}},\
  }\bibfield  {title} {\bibinfo {title} {Purely hydrodynamic origin for
  swarming of swimming particles},\ }\href@noop {} {\bibfield  {journal}
  {\bibinfo  {journal} {Physical Review E}\ }\textbf {\bibinfo {volume} {93}},\
  \bibinfo {pages} {043114} (\bibinfo {year} {2016})}\BibitemShut {NoStop}%
\bibitem [{\citenamefont {Bleibel}\ \emph {et~al.}(2014)\citenamefont
  {Bleibel}, \citenamefont {Dom{\'\i}nguez}, \citenamefont {G{\"u}nther},
  \citenamefont {Harting},\ and\ \citenamefont
  {Oettel}}]{bleibel2014hydrodynamic}%
  \BibitemOpen
  \bibfield  {author} {\bibinfo {author} {\bibfnamefont {J.}~\bibnamefont
  {Bleibel}}, \bibinfo {author} {\bibfnamefont {A.}~\bibnamefont
  {Dom{\'\i}nguez}}, \bibinfo {author} {\bibfnamefont {F.}~\bibnamefont
  {G{\"u}nther}}, \bibinfo {author} {\bibfnamefont {J.}~\bibnamefont
  {Harting}},\ and\ \bibinfo {author} {\bibfnamefont {M.}~\bibnamefont
  {Oettel}},\ }\bibfield  {title} {\bibinfo {title} {Hydrodynamic interactions
  induce anomalous diffusion under partial confinement},\ }\href@noop {}
  {\bibfield  {journal} {\bibinfo  {journal} {Soft Matter}\ }\textbf {\bibinfo
  {volume} {10}},\ \bibinfo {pages} {2945} (\bibinfo {year}
  {2014})}\BibitemShut {NoStop}%
\bibitem [{\citenamefont {Blake}(1971{\natexlab{b}})}]{blake1971note}%
  \BibitemOpen
  \bibfield  {author} {\bibinfo {author} {\bibfnamefont {J.~R.}\ \bibnamefont
  {Blake}},\ }\bibfield  {title} {\bibinfo {title} {A note on the image system
  for a stokeslet in a no-slip boundary},\ }in\ \href@noop {} {\emph {\bibinfo
  {booktitle} {Mathematical Proceedings of the Cambridge Philosophical
  Society}}},\ Vol.~\bibinfo {volume} {70}\ (\bibinfo {organization} {Cambridge
  University Press},\ \bibinfo {year} {1971})\ pp.\ \bibinfo {pages}
  {303--310}\BibitemShut {NoStop}%
\bibitem [{\citenamefont {Mitchell}\ and\ \citenamefont
  {Spagnolie}(2015)}]{mitchell2015sedimentation}%
  \BibitemOpen
  \bibfield  {author} {\bibinfo {author} {\bibfnamefont {W.~H.}\ \bibnamefont
  {Mitchell}}\ and\ \bibinfo {author} {\bibfnamefont {S.~E.}\ \bibnamefont
  {Spagnolie}},\ }\bibfield  {title} {\bibinfo {title} {Sedimentation of
  spheroidal bodies near walls in viscous fluids: glancing, reversing, tumbling
  and sliding},\ }\href@noop {} {\bibfield  {journal} {\bibinfo  {journal}
  {Journal of Fluid Mechanics}\ }\textbf {\bibinfo {volume} {772}},\ \bibinfo
  {pages} {600} (\bibinfo {year} {2015})}\BibitemShut {NoStop}%
\bibitem [{\citenamefont {de~Graaf}\ and\ \citenamefont
  {Stenhammar}(2017)}]{de2017stirring}%
  \BibitemOpen
  \bibfield  {author} {\bibinfo {author} {\bibfnamefont {J.}~\bibnamefont
  {de~Graaf}}\ and\ \bibinfo {author} {\bibfnamefont {J.}~\bibnamefont
  {Stenhammar}},\ }\bibfield  {title} {\bibinfo {title} {Stirring by periodic
  arrays of microswimmers},\ }\href@noop {} {\bibfield  {journal} {\bibinfo
  {journal} {Journal of Fluid Mechanics}\ }\textbf {\bibinfo {volume} {811}},\
  \bibinfo {pages} {487} (\bibinfo {year} {2017})}\BibitemShut {NoStop}%
\end{thebibliography}%

\end{document}


\setcounter{secnumdepth}{2}

\renewcommand{\thefigure}{S\arabic{figure}}

\title{Supplementary Material: Clustering and emergent hyperuniformity by breaking microswimmer shape and actuation symmetries}
\author{Anson G. Thambi}
\affiliation{Department of Mechanical Engineering, University of Hawai’i at 
M{\=a}noa, 2540 Dole Street, Holmes Hall 302, Honolulu, Hawaii 96822, USA}

\author{William E. Uspal}
\email{uspal@hawaii.edu}
\affiliation{Department of Mechanical Engineering, University of Hawai’i at 
M{\=a}noa, 2540 Dole Street, Holmes Hall 302, Honolulu, Hawaii 96822, USA}
\affiliation{International Institute for Sustainability with Knotted Chiral Meta Matter (WPI-SKCM\textsuperscript{2}),
1-3-1 Kagamiyama, Higashi-Hiroshima, Hiroshima 739-8526, Japan}

\date{\today}
\maketitle

\section{ Symmetry and Flow Behavior with Varying $\delta$}

\begin{figure}[h!bt]
    \centering
    \includegraphics[width=\textwidth]{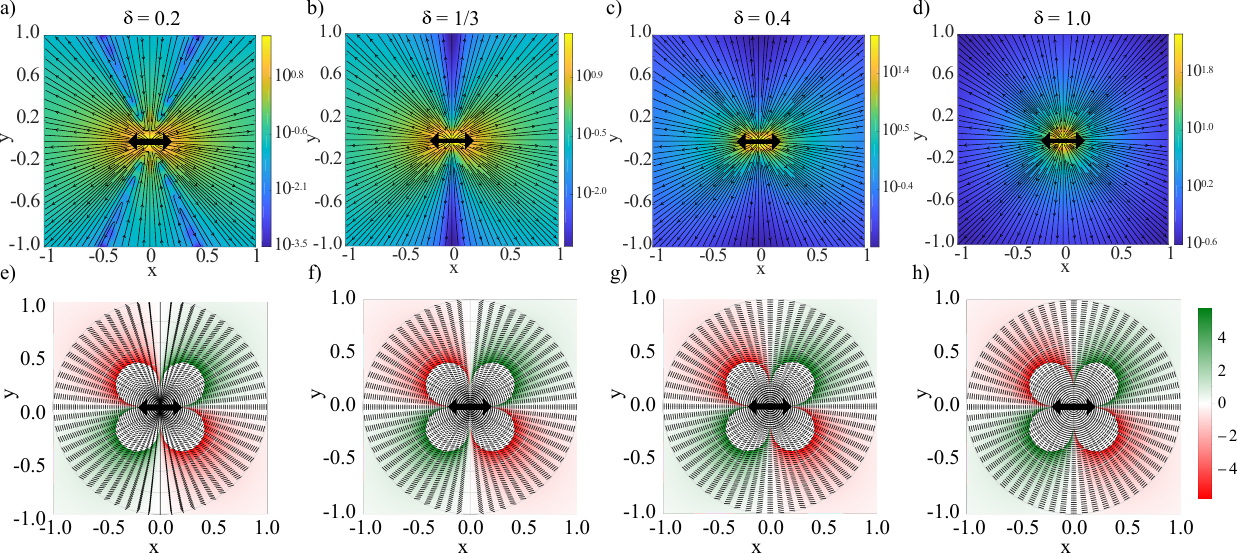}
    \caption{\label{fig:flowfields}(a)--(d) Flow fields generated by a non-axisymmetric pusher-type stresslet located at the origin \((0,0)\) for asymmetry parameters \(\delta = 0.2, \frac{1}{3}, 0.4, 1.0\). The stresslet, oriented along the \(\hat{x}\)-direction, has strength \(\sigma_0/(\mu L_0^2 U_0) = -6\) and is depicted by a double-headed arrow. The colormap indicates the dimensionless velocity magnitude $|\mathbf{u}|/U_0$.
(e)--(h) Corresponding vorticity fields (colormap, in units of $U_0/L_0$) and elongation axes of the rate-of-strain tensor (black lines) for the same stresslet configurations. The black lines represent the local orientation of the elongation axis across the domain.}
\end{figure}
As discussed in the main text, introducing the asymmetry parameter \(\delta\), which quantifies the deviation of the stresslet from its axisymmetric form, still preserves two planes of mirror symmetry in the generated flow field, as illustrated in Fig.~\ref{fig:flowfields}. Notably, at \(\delta = \frac{1}{3}\) (Fig.~\ref{fig:flowfields}b), the \(\mathbf{\hat{c}}\mathbf{\hat{c}}\) component of the stresslet cancels out completely, resulting in the absence of any inflow regions in the velocity field. This is in contrast to the case \(\delta < \frac{1}{3}\), where inflow is present (see Fig.~\ref{fig:flowfields}a). For values \(\delta > \frac{1}{3}\) (Figs.~\ref{fig:flowfields}c and d), the flow becomes purely outward across the domain.

\FloatBarrier

\section{Variation of Inter-Particle Distance with Cluster Size}

\begin{figure}[h!bt]
    \centering
    \includegraphics[scale=0.86]{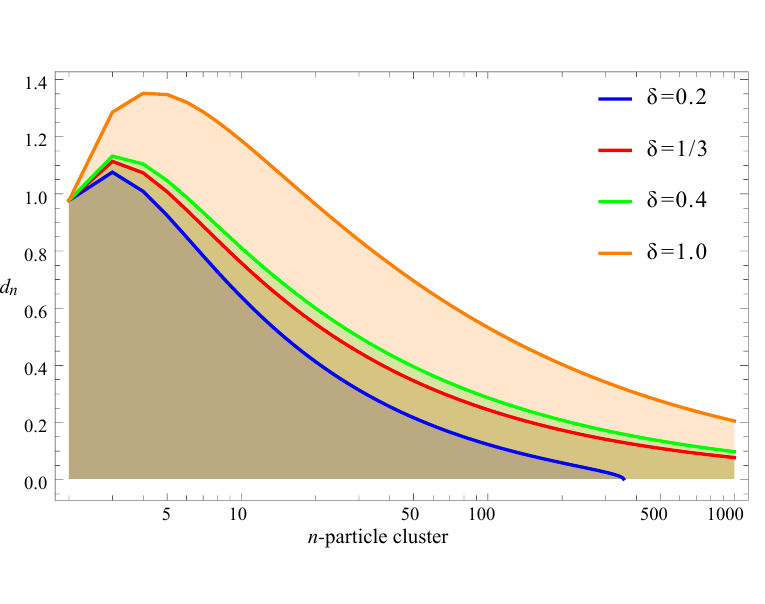}
    \caption{\label{fig:dn}Plot of the inter-particle distance \(d_n\) (in units of $L_0$), computed using Eq.~(7 ) from the main text, as a function of the number of particles \(n\) in a cluster. Each cluster corresponds to a regular \(n\)-gon configuration, where \(d_n\) represents the distance between adjacent particles. For example, \(n = 4\) corresponds to a square configuration. The stresslet strength is fixed at \(\sigma_0/(\mu L_0^2 U_0) = -6\) with $U_s/U_0 = 0.5$, and results are shown for values of the asymmetry parameter \(\delta = 0.2,\, \frac{1}{3},\, 0.4,\, 1.0\). We note that for $\delta = 0.2$, there is no real-valued solution for $d_n$ at large $n$ for pushers ($\sigma_0 < 0)$, although a real-valued solution emerges for pullers ($\sigma_0 > 0$; not shown here). This transition arises from the requirement that the cluster-sourced flow must be radially outward at the particles' positions.} 
\end{figure}

\FloatBarrier

\section{Details of the simulation method}

In the simulations, the semi-major axis is used as the characteristic length scale $L_0$. The characteristic velocity scale $U_0$ is arbitrary, and a characteristic force can be obtained as $F_0 = \mu L_0^2 / T_0$. We also define a characteristic time $T_0 \equiv L_0 / U_0$.

As discussed in the main text, the equations of motion can be extended to include an excluded volume force:
\begin{equation}
\textbf{U}^i = U_s^i   \mathbf{\hat{d}}^i + \textbf{u}(\textbf{x}_i) + \mu_t \mathbf{F}_{\textrm{ev},i}(\{\mathbf{x}_j, \mathbf{d}_j\}),
\end{equation}
where the excluded volume force is a function of the positions $\mathbf{x}_j$ and orientations $\mathbf{d}_j$ of all $N$ particles. We neglect the configuration dependence of the translational hydrodynamic mobility $\mu_t$. The practical importance of the excluded volume force is to prevent  particles from approaching too closely during the time evolution of the system, since the expression used to determine $\mathbf{u}(\mathbf{x}_i)$ is singular in the interparticle distance. We note that the excluded volume force is not included in, or necessary to, the prediction of multi-particle stable states in the kinetic theory of the main text.

In order to compute the excluded volume force, we use a ``shish-kebab'' model in which each particle is associated with multi-bead rod. Each rod has $2N_b + 1$ beads. The beads are spherical (but may have different radii $R_{\alpha}$), and the multi-bead rod translates and co-rotates with the particle.  The excluded volume force on particle $i$ is the sum of forces on beads $\alpha$ associated with particle $i$:
\begin{equation}
\mathbf{F}_{\textrm{ev},i} = \sum_{\alpha = -N_b}^{N_b} \mathbf{F}_{\textrm{b},i,\alpha}.
\end{equation}
The force $\mathbf{F}_{\textrm{b},i,\alpha}$ is computed from pairwise interactions of with beads $\beta$ on all particles $j \neq i$:
\begin{equation}
\mathbf{F}_{\textrm{b},i,\alpha} = \sum_{j \neq i}^{N} \sum_{\beta=-N_b}^{N_b} \mathbf{F}_{\textrm{bb}}(\mathbf{x}_{i,\alpha}, \mathbf{x}_{j,\beta}).
\end{equation}
Here $\mathbf{x}_{i,\alpha}$ indicates the position of bead $\alpha$ on particle $i$. The bead-bead  force is given by a soft repulsion
\begin{equation}
\label{eq:bead_bead_soft_ev}
\mathbf{F}_{\textrm{bb}}(\mathbf{x}_{i,\alpha}, \mathbf{x}_{j,\beta}) = -k(|\mathbf{x}_{j,\beta} - \mathbf{x}_{i,\alpha}|-R_{i,\alpha} - R_{j,\beta}) \frac{\mathbf{x}_{j,\beta} - \mathbf{x}_{i,\alpha}}{|\mathbf{x}_{j,\beta} - \mathbf{x}_{i,\alpha}|},
\end{equation}
where $R_{j,\beta}$ and $R_{i,\alpha}$ are the radii of beads $\beta$ and $\alpha$ on particles $j$ and $i$, respectively.

Likewise, the equation for the  time evolution of the particle orientation can be extended as 
\begin{equation}
\mathbf{\dot{\hat{d}}}^i = (\bm{\mathcal{I}}-\mathbf{\hat{d}}^i \mathbf{\hat{d}}^i)\cdot (\Gamma_i \textbf{E}(\mathbf{x}_i)+\textbf{W}(\mathbf{x}_i))\cdot\mathbf{\hat{d}}^i + \mu_r \sum_{\alpha=-N_b}^{N_b}(\mathbf{x}_{i,\alpha} - \mathbf{x}_i) \times \mathbf{F}_{b,i,\alpha},
\label{eq:jeffery_extended}
\end{equation}
where we neglect the configuration dependence of the rotational hydrodynamic mobility $\mu_r$.

To parameterize the soft excluded volume force, we choose $\mu_t k L_0/U_0 = 1$. Each particle has an identical number of beads $N_b = \textrm{ceil}((r_e-1)/2)$ and bead radii $R_{\alpha} = 2/r_e$, where $\textrm{ceil}$ is the ceiling function. We specify the bead locations by considering a ``primed'' frame in which the particle center is located at $(0,0)$ and the semi-major axis is in the $y'$ direction. In this frame, bead locations are $x_{\alpha}' = 0$ and $y_{\pm \alpha}' = \pm 2 \alpha/r_e$.  We also choose $\mu_r L_0^2 = \mu_t$ for simplicity.

For certain simulation parameters $\Gamma$ and $\delta$, the soft excluded volume force given in Eq. \ref{eq:bead_bead_soft_ev} does not adequately prevent close approach of particles. In these situations, we use the Heyes-Melrose algorithm, which explicitly imposes a ``hard'' excluded volume constraint between beads \cite{heyes1993brownian}. Specifically, when beads associated with two different particles overlap, we displace both particles along the line connecting the centers of the two beads. The displacement is determined to put the two beads just out of contact.

For simulations of collective dynamics, we use a simple realization of periodic boundary conditions. The minimum image convention is employed to compute interactions between particles under periodic boundary conditions, wherein each particle interacts only with the nearest periodic image of every other particle, rather than summing interactions over all periodic copies \cite{frenkel2002understanding}.

We fix $U_s/U_0 = 0.5$ for all particles. The equations of motion are integrated numerically using the Euler algorithm with timestep $\delta/T_0 = 0.05$.

\section{Linear Stability Analysis}

The linear stability of fixed points corresponding to particle positions arranged in \(n\)-particle clusters is examined, where the particles are configured in a regular polygonal arrangement. This analysis is carried out by evaluating the Jacobian matrix of the system and examining its eigenvalues. Due to the analytical complexity of this procedure, the Jacobian is computed numerically using finite difference methods.

The Jacobian matrix \(\mathbf{J}\) for a \(n\)-gon configuration is constructed by considering the linearized particle dynamics for small perturbations away from a fixed point, using the expressions for translational and rotational motion given in Eqs.~(3) and (4) of the main text.

\begin{equation}
\mathbf{J} =
\begin{pmatrix}
\frac{\partial u_x^{(1)}}{\partial x_1} & \frac{\partial u_x^{(1)}}{\partial y_1} & \frac{\partial u_x^{(1)}}{\partial \hat{\mathbf{d}}_1} & \cdots & \frac{\partial u_x^{(1)}}{\partial x_n} & \frac{\partial u_x^{(1)}}{\partial y_n} & \frac{\partial u_x^{(1)}}{\partial \hat{\mathbf{d}}_n} \\
\frac{\partial u_y^{(1)}}{\partial x_1} & \frac{\partial u_y^{(1)}}{\partial y_1} & \frac{\partial u_y^{(1)}}{\partial \hat{\mathbf{d}}_1} & \cdots & \frac{\partial u_y^{(1)}}{\partial x_n} & \frac{\partial u_y^{(1)}}{\partial y_n} & \frac{\partial u_y^{(1)}}{\partial \hat{\mathbf{d}}_n} \\
\frac{\partial \dot{\hat{\mathbf{d}}}_1}{\partial x_1} & \frac{\partial \dot{\hat{\mathbf{d}}}_1}{\partial y_1} & \frac{\partial \dot{\hat{\mathbf{d}}}_1}{\partial \hat{\mathbf{d}}_1} & \cdots & \frac{\partial \dot{\hat{\mathbf{d}}}_1}{\partial x_n} & \frac{\partial \dot{\hat{\mathbf{d}}}_1}{\partial y_n} & \frac{\partial \dot{\hat{\mathbf{d}}}_1}{\partial \hat{\mathbf{d}}_n} \\
\vdots & \vdots & \vdots & \ddots & \vdots & \vdots & \vdots \\
\frac{\partial \dot{\hat{\mathbf{d}}}_n}{\partial x_1} & \frac{\partial \dot{\hat{\mathbf{d}}}_n}{\partial y_1} & \frac{\partial \dot{\hat{\mathbf{d}}}_n}{\partial \hat{\mathbf{d}}_1} & \cdots & \frac{\partial \dot{\hat{\mathbf{d}}}_n}{\partial x_n} & \frac{\partial \dot{\hat{\mathbf{d}}}_n}{\partial y_n} & \frac{\partial \dot{\hat{\mathbf{d}}}_n}{\partial \hat{\mathbf{d}}_n}
\end{pmatrix}.
\end{equation}
The Jacobian is evaluated at the symmetric configuration where particles are arranged in a regular \(n\)-gon centered at the origin. Since \(d_n\) (Eqn. (5)) denotes the distance between adjacent particles, then the radius \(R\) of the circumscribed circle is:
\begin{equation}
R = \frac{d_n}{2 \sin\left( \frac{\pi}{n} \right)}.
\end{equation}
The position vector of the \(l\)-th particle \((l = 1, 2, \dots, n)\) is given by:
\begin{equation}
\mathbf{x}_l = 
R \begin{pmatrix}
\cos\left( \frac{2\pi (l - 1)}{n} \right) \\
\sin\left( \frac{2\pi (l - 1)}{n} \right) \\
0
\end{pmatrix}.
\end{equation}
The corresponding orientation angle is:
\begin{equation}
\theta_l = \frac{2\pi(l - 1)}{n} + \pi,
\end{equation}
and the orientation vector of the \(l\)-th particle is:
\begin{equation}
\hat{\mathbf{d}}_l =
\begin{pmatrix}
\cos \theta_l \\
\sin \theta_l \\
0
\end{pmatrix}.
\end{equation}
The stability results obtained using the Jacobian for different $n$-gon configuration of the particle clusters as a function of $\delta$ and $\Gamma$ are plotted in the Fig. 3c in the main text.

\section{Squirmer model for oblate spheroids}

We consider $\mathcal{N}$ oblate spheroidal particles in an unbounded and incompressible Newtonian solution. The fluid velocity $\mathbf{u}(\mathbf{x})$ is governed by the Stokes equation $-\nabla p + \mu \nabla^2 \mathbf{u} = 0$, where $\mu$ is the dynamic viscosity of the solution, and $p(\mathbf{x})$ is the pressure, as well as the incompressibility condition $\nabla \cdot \mathbf{u} = 0$. On the surface of particle $i \in \{1, 2, \ldots \mathcal{N}\}$, the velocity obeys the boundary condition $\mathbf{u} = \mathbf{U}^i + \bm{\Omega}^i \times (\mathbf{x} - \mathbf{x}_i) + \mathbf{v}_{s,i}(\mathbf{x})$. Here, $\mathbf{x}_i$ is the geometric centroid of particle $i$; $\mathbf{U}^i$ and $\bm{\Omega}^i$ are the translational and angular velocities, respectively, of particle $i$; and $\mathbf{v}_{s,i}(\mathbf{x})$ is the so-called ``slip velocity'' at the surface of particle $i$. The slip velocity represents the interfacial actuation that drives the swimming motion of a squirmer, and is specified below. Additionally, far away from the particles, the velocity vanishes, i.e., $|\mathbf{u}(\mathbf{x})| \rightarrow 0$ as $|\mathbf{x}| \rightarrow \infty$. Each particle is force-free and torque-free, giving $\int_{S_i} \bm{\sigma} \cdot \mathbf{\hat{n}} \, dS = 0$ and $\int_{S_i} (\mathbf{x} -\mathbf{x}_i) \times \bm{\sigma} \cdot \mathbf{\hat{n}} \, dS = 0$, where $S_i$ is the surface of particle $i$, and the normal vector $\mathbf{\hat{n}}$ is defined to point into the fluid. The force-free and torque-free conditions close the system of equations for the $6 \mathcal{N}$ unknown components of $\mathbf{U}^i$ and $\bm{\Omega}^i$.

We specify $\mathbf{v}_{s,i}(\mathbf{x})$ as the superposition of so-called ``squirming modes'' in a frame that is translating and rotating with the particle. If the propulsion axis of particle $i$ is aligned with the vector $\mathbf{\hat{d}}^i$, we write:
\begin{equation}
\mathbf{v}_{s,i}(\mathbf{x}) = \mathbf{v}_{\textrm{ax},i}(\mathbf{x}) + \mathbf{v}_{\textrm{nax},i}(\mathbf{x}). 
\end{equation}
Here, we distinguish axisymmetric and non-axisymmetric contributions. For the axisymmetric contribution, we use the modes developed for oblate spherodial coordinates in Ref. \citenum{poehnl2020axisymmetric}. We briefly summarize these modes as developed for a single particle; the generalization to multiple particles is straightforward. The axisymmetric part of the slip is written as:  
\begin{equation}
\label{axi_mode}
\mathbf{v}_{\textrm{ax}}(\mathbf{x}) = \lambda_0(\lambda_0^2 + \xi^2)^{-1/2} \sum_{n=1} B_n P_n^1(\xi) \; \mathbf{\hat{e}}_{\xi}, 
\end{equation}
where $B_n$ is the amplitude of the axisymmetric squirming mode of order $n$, $P_n^1$ is an associated Legendre polynomial, and $(\lambda, \xi, \varphi)$ are oblate spheroidal coordinates. The origin of the coordinate system is at the geometric centroid of the particle, and the coordinate system is co-rotating and co-moving with the particle. In this coordinate system, the surface of the particle is defined to be $\lambda = \lambda_0$, and the vector $\mathbf{\hat{e}}_{\xi}$ is tangent to the surface of the particle in the polar direction. The squirming mode amplitudes $B_n$ are constants (with dimensions of velocity) and specified by fiat. Here, we choose to truncate the axisymmetric squirming modes to $B_1$, which gives self-propulsion, and $B_2$, which contributes to the axisymmetric part of the stresslet tensor. Further details are in Ref. \citenum{poehnl2020axisymmetric}.

For the non-axisymmetric part of the slip velocity, we use the mode developed in Ref. \citenum{poehnl2023shape}. This mode is partially given in terms of a Cartesian coordinate system $(x',y',z')$ that has its origin at the geometric centroid of the particle  is co-rotating and co-moving with the particle. It is written as follows:
\begin{equation}
\label{eq:nasym_slip}
\mathbf{v}_{\textrm{nax}}(\mathbf{x}) = \tilde{B} [\text{sign}(z')(\cos \varphi \, \mathbf{\hat{e}}_\xi-\mathbf{\hat{e}}_\varphi(\mathbf{\hat{e}}_\xi\cdot\mathbf{\hat{e}}_{{z'}}))
 -\text{sign}(x')(\sin \varphi \, \mathbf{\hat{e}}_\xi -\hat{e}_\varphi (\mathbf{\hat{e}}_\xi \cdot \mathbf{\hat{e}}_{x'})) ] \; H(y'),
\end{equation}
where $H(y')$ is the step function, and $\mathbf{\hat{e}}_\varphi$ is the azimuthal tangent vector. The quantity $\tilde{B}$ is the non-axisymmetric squirming mode amplitude. Like the axisymmetric mode amplitudes, it is constant and specified by fiat. This mode determines the non-axisymmetric part of the stresslet tensor. The quantity $\tilde{B}$ is proportional to $\delta \sigma_0$, and the constant of proportionality depends on the particle aspect ratio, as detailed in Ref. \citenum{poehnl2023shape}.

\section{Comparison of squirmer model and kinetic theory}

\begin{figure}[h!tb]
    \centering
    \includegraphics[scale=0.8]{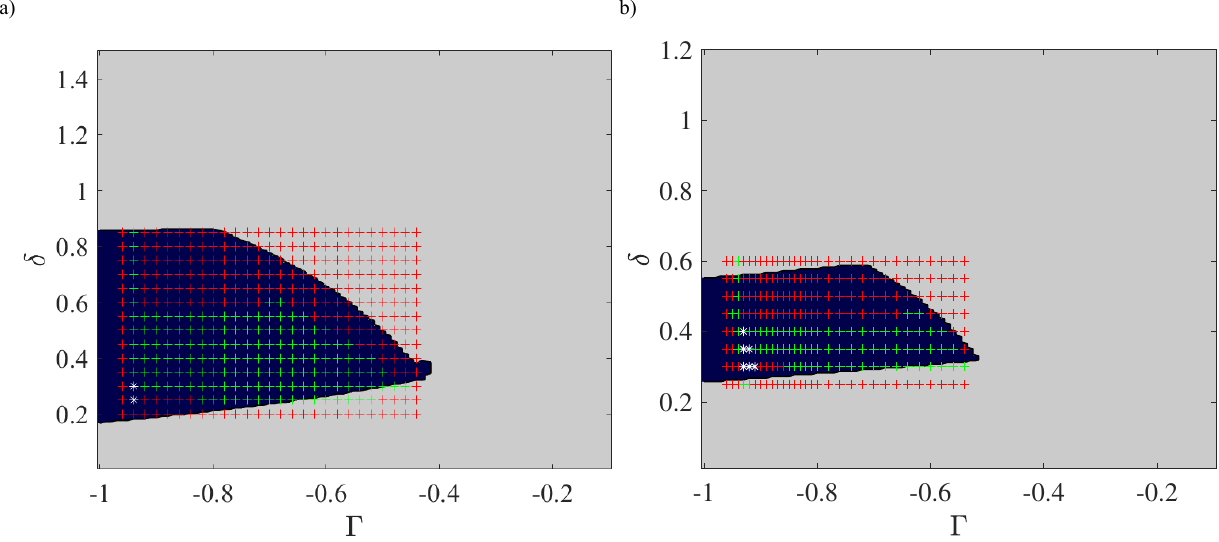}
    \caption{\label{fig:bem_vs_ppms} Phase maps for (a) 3-particle and (b) 4-particle clusters. The background shading represents the stability predictions from kinetic theory, which are independent of the stresslet strength $\sigma_0$. 
Symbols correspond to results from numerical simulations of non-axisymmetric squirmers (Eqs.~\ref{axi_mode} and~\ref{eq:nasym_slip}) with fixed $B_1 = 0.3\, U_0$, and $B_2/U_0$ varied within a range of $-0.25$ to $-1.10$ to maintain a constant stresslet strength $\sigma_0/(\mu L_0^2 U_0) \approx -60$, across different values of $\Gamma$ and $\tilde{B}$. 
Green plus signs mark conditions under which stable regular polygonal clusters are observed in the squirmer model. White stars indicate stable clusters that do not form regular polygons. For 3-particle clusters, this deviation results in a net velocity, making them motile.}
\end{figure}

\FloatBarrier

\FloatBarrier
\section{Mean Speed Dynamics Outside the Absorbing State}

\begin{figure}[h!tb]
    \centering
    \includegraphics[scale=0.5]{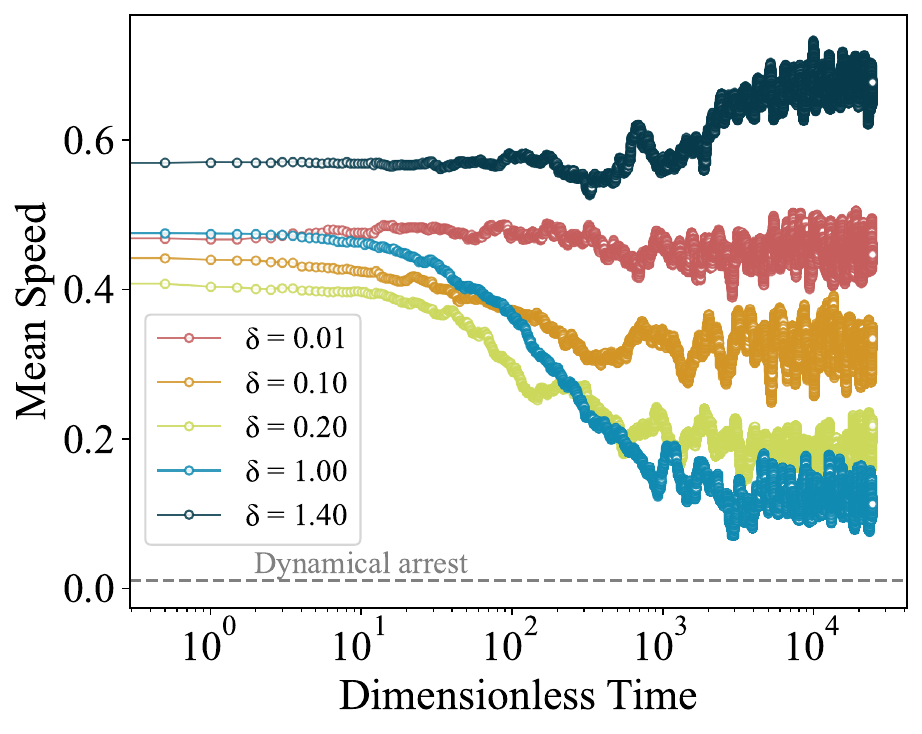}
    \caption{\label{fig:mean_speed} Time evolution of the mean speed (in units of $U_0$) for various $\delta$ values that lie outside the absorbing state regime and therefore do not lead to dynamical arrest. As evident from the plot, the mean speed in each case reaches a finite asymptotic value and fluctuates around it, rather than decaying toward the dynamical arrest threshold. The grey horizontal line at velocity $\approx 0.01$ indicates the characteristic arrest speed identified in our system, which is not approached by the trajectories shown here.}
\end{figure}

\FloatBarrier

\section{Quantifying Flocking via Polar Order}

\begin{figure}[h!tb]
    \centering
    \includegraphics[scale=0.77]{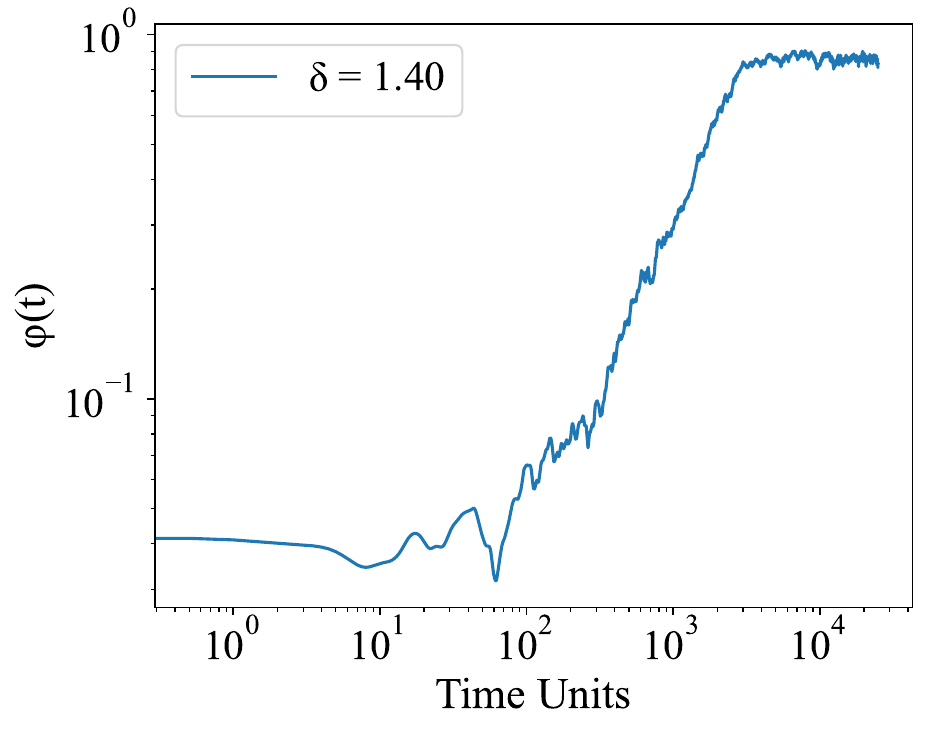}
   \caption{\label{fig:polar}
Time evolution of the polar order parameter $\varphi$ for the flocking state at $\delta = 1.4$. The plot demonstrates the emergence of flocking behavior, as $\varphi$ steadily increases and saturates near unity, indicating strong global alignment among the particles.}
\end{figure}
    
To characterize the extent of flocking in the system,we define the polar order parameter:
\begin{equation}
\varphi = \frac{1}{N} \left| \sum_{i=1}^N \hat{\mathbf{d}}^i \right|.
\end{equation}
where $\hat{\mathbf{d}}^i$ is the orientation of the $i$-th particle and $N$ is the total number of particles. This parameter measures the global alignment of orientations.

In a disordered state, particles move randomly with no preferred direction, resulting in negligible net motion and hence $\varphi \to 0$. Conversely, when the system undergoes flocking, particles align and travel collectively in a spontaneously selected direction, leading to $\varphi \to 1$ as can be seen from the Fig \ref{fig:polar}.

\FloatBarrier

\section{Hyperuniformity at different number Densities}

\begin{table}[h!bt]
\centering
\caption{Fitted power-law exponents \(\alpha\) from the small-\(q\) behavior of the structure factor, \(S(q) \sim q^\alpha\), at different number densities for various values of \(\delta\).}
\label{tab:alpha_fits}
\begin{tabular}{c|c|c|c|c|c}
\hline
\(\phi\) & $\delta = 0.800$ & $0.530$ & $0.475$ & $0.445$ & $0.333$ \\
\hline
$0.004$ & $5.8884 \pm 0.1122$& $6.4279 \pm 0.0952$& $6.4201 \pm 0.1294$& $6.0776 \pm 0.1254$& $4.8985\pm 0.1305$\\
$0.005$ & $6.1411 \pm 0.0902$ & $6.690 \pm 0.1376$ & $6.3213 \pm 0.0953$ & $6.0701 \pm 0.1045$ & $4.9627 \pm 0.1046$ \\
$0.006$ & $6.4099 \pm 0.0778$ & $6.0576 \pm 0.0875$ & $6.2833 \pm 0.0717$ & $5.8957 \pm 0.0797$ & $4.7244 \pm 0.0747$ \\
\hline
\end{tabular}
\end{table}

\begin{figure}[h!bt]
    \centering
    \includegraphics[width=\textwidth]{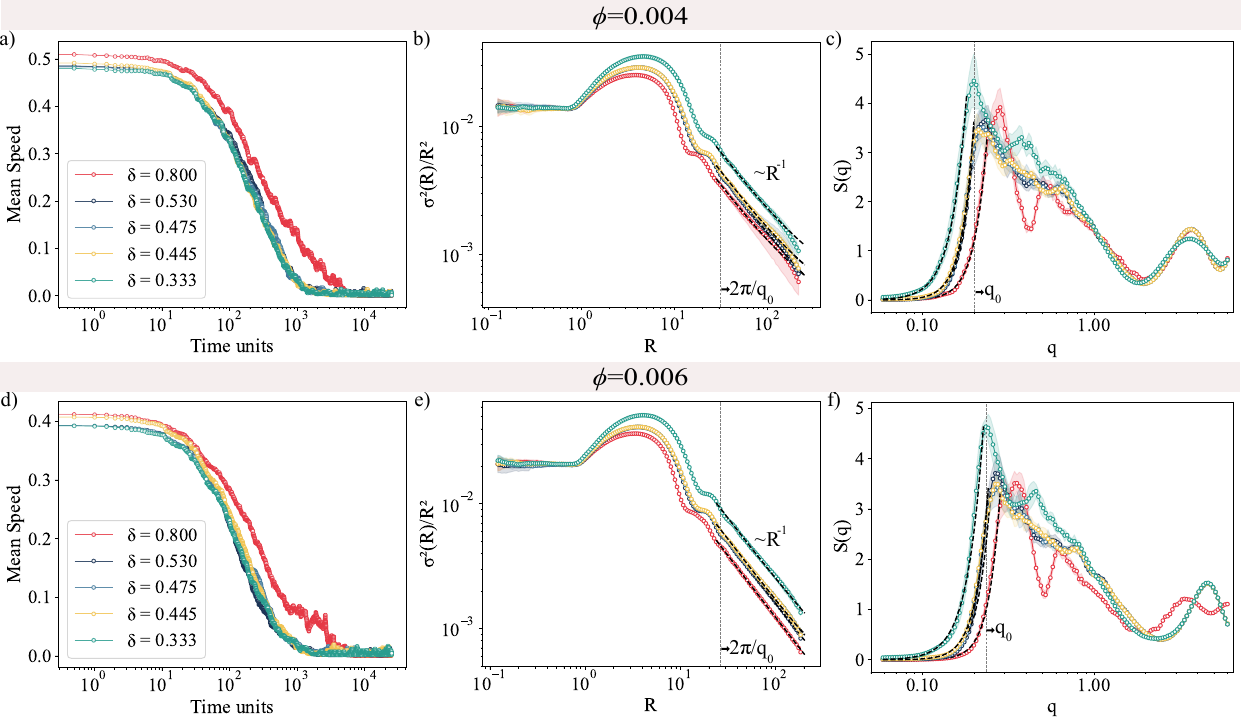}
   \caption{\label{fig:phi_density}
(a–c) Results for number density \(\phi = 0.004\), and (d–f) corresponding results for \(\phi = 0.006\), both with stresslet strength \(\sigma_0/\mu L_0^2 U_0 = -14.95\). 
(a, d) Time evolution of the mean particle speed (in units of \(U_0\)) for selected values of \(\delta\). In each case, the mean speed decays below 0.01 within \(10^4\) time units, indicating a transition to a dynamically arrested absorbing state.
(b, e) Scaled number variance \(\sigma^2(R)/R^2\) as a function of window radius \(R\) (in units of \(L_0\)), showing a \(1/R\) decay at large \(R\) (black dashed line), characteristic of Class I hyperuniformity. The vertical grey dashed line marks the crossover length scale \(2\pi/q_0\) beyond which this scaling emerges.
(c, f) Structure factor \(S(q)\) plotted against wavevector \(q\) (in units of \(1/L_0\)). The small-\(q\) behavior follows a power-law decay \(S(q) \sim q^{\alpha}\) (black dashed line), consistent with Class I hyperuniformity. Exact values of the exponent \(\alpha\) are provided in Table~1. The vertical grey dashed line indicates the first peak at \(q_0\). These results complement those at \(\phi = 0.005\) in the main text (Fig.~5), confirming the robustness of hyperuniformity across densities.
}

\end{figure}

\FloatBarrier

\section{Calculation of the static structure factor}

\begin{figure}[h!tb]
    \centering
    \includegraphics[scale=0.75]{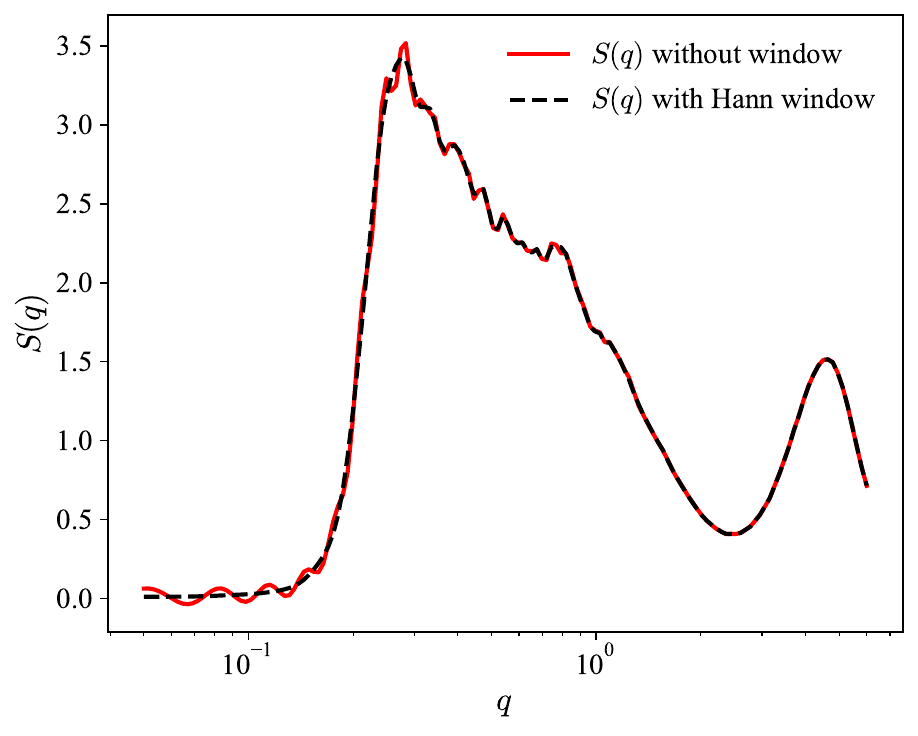}
   \caption{\label{fig:hann}Comparison of the structure factor \( S(q) \) calculated without and with the application of a Hann window using eqns. \ref{eqn:S(q)_2d} and \ref{Sq_hann} respectively for $\delta = 0.4750$ at $\phi=0.006$. The plot shows that applying the Hann window significantly suppresses the artificial oscillations (ripples) at low wavevectors \( q \).}
\end{figure}

For an isotropic system in $d$ dimensions, the static structure factor $S(q)$ can be expressed analytically as a Hankel transform of the pair correlation function $g(r)$ as follows \cite{hawat2023estimating}:
\begin{equation}
\label{eqn:s(q)_D}
    S(q) \;=\; 1 \;+\; \phi\,(2\pi)^{d/2}\,q^{\,1-d/2}\int_{0}^{\infty}\![\,g(r)-1\,]\,r^{d/2}\,J_{d/2-1}(q\,r)\,dr \,
\end{equation}
Here $J_\nu(x)$ is the Bessel function of the first kind of order $\nu$. Hence for a $2$-dimensional system, this equation reduces to the following form:
\begin{equation}
    \label{eqn:S(q)_2d}
    S(q) = 1\;+\;2\pi\phi\int_0^\infty[g(r)-1]J_0(qr)\;r\;dr
\end{equation}
A known challenge in using this equation is that it produces artificial oscillations at low $q$-values due to the truncation of the integral at $r_{max}$ which is half the box size \cite{soper2012use,cheng2003fourier}. However, this can be mitigated by using a window function such as the Hanning window \cite{press1992numerical,morris1996spatio}. We therefore use the Hann function as follows:

\begin{equation}
\label{hann}
    W(r) =
\begin{cases}
1, & \text{if } r <\dfrac{r_{max}}{2} \\[8pt]
\frac{1}{2} \left[ 1 + \cos\left( 2\pi \left( \dfrac{r}{r_{\max}} - \dfrac{1}{2} \right) \right) \right], & \text{if } \dfrac{r_{max}}{2} \leq r \leq r_{max} \\[8pt]
0, & \text{if } r > r_{max}
\end{cases}
\end{equation}
Hence the $S(q)$ with the window is as follows:
\begin{equation}
\label{Sq_hann}
   S(q) = 1\;+\;2\pi\phi\int_0 ^{r_{max}}[g(r)-1]J_0(qr) \;r\;W(r)\;dr
\end{equation}
This equation was used for the structure factor calculation, and Fig. \ref{fig:hann} exhibits the suppression of ripple artifact caused by the truncation once windowing is enabled.

\FloatBarrier
\section{List of videos}
The following videos correspond to different dynamical states observed in the system at a number density of $\phi = 0.005$. Particle sizes have been enlarged for visual clarity.

\begin{itemize}
    \item \textbf{Video V1:} Flocking, $\delta = 1.4$
    \item \textbf{Video V2:} Stable motile clusters, $\delta = 1$
    \item \textbf{Videos V3 and V4:} Absorbing states, $\delta =0.800$ and $\delta = 0.333 $
    \item \textbf{Video V5:} Unstable motile clusters, $\delta = 0.2$
\end{itemize}

\bibliography{SM_bib}